\documentclass[showpacs,amssymb,twocolumn,floatfix,pre,aps,notitlepage]{revtex4-1}

\usepackage{amssymb,amsfonts,amsmath,bm}
\usepackage{graphics,graphicx}
\usepackage{color}

\newcommand{\dhc}{d_{\rm hc}}

\newcommand{\remma}[1]{{\color{black} #1}}
\begin{document}

\title{Strong-coupling theory of counterions with hard cores
between symmetrically charged walls}

\author{Ladislav \v{S}amaj$^{1}$} 
\author{Martin Trulsson$^2$}
\author{Emmanuel Trizac$^3$}

\affiliation{
$^1$Institute of Physics, Slovak Academy of Sciences, Bratislava, Slovakia \\
$^2$Theoretical Chemistry, Lund University, Sweden \\
$^3$Universit\'e Paris-Saclay, CNRS, LPTMS, 91405, Orsay, France. 
}

\date{\today} 

\begin{abstract}
By a combination of Monte  Carlo simulations and analytical calculations, we investigate the effective interactions between highly charged planar 
interfaces, neutralized by mobile counterions \remma{(salt-free system)}. While most previous analysis have focused on point-like counterions, we treat them as charged hard spheres.
We thus work out the fate of like-charge attraction when steric effects are at work. The analytical approach partitions counterions in two sub-populations, one for each plate,
and integrates out one sub-population to derive an effective Hamiltonian 
for the remaining one. The effective Hamiltonian features plaquette four-particle interactions, and 
it is worked out by computing a Gibbs-Bogoliubov 
inequality for the free energy. At the root of the treatment is the fact that 
under strong electrostatic coupling, the system of charges forms an ordered arrangement,
that can be affected by steric interactions. Fluctuations around the reference positions are accounted for. To dominant order at high coupling, it is found that steric effects do not significantly affect the interplate effective pressure,
apart at small distances where hard sphere overlap are unavoidable, and thus rule out configurations. 
\end{abstract}

\maketitle

\renewcommand{\theequation}{1.\arabic{equation}}
\setcounter{equation}{0}

\section{Introduction} \label{Sec.intro}

The dominant part of colloids release micro-ions of low valence from the surfaces at deionized conditions
\cite{Levin02,Andelman06,Palberg04}. These mobile so-called counterions can be regarded as identical classical particles 
interacting via the three-dimensional $1/r$ Coulomb potential.
The charged surface with the surrounding counterions form 
in thermal equilibrium a neutral electric double layer
\cite{Attard96,Hansen00,Messina09}.
The geometry of two parallel similarly and uniformly charged walls at distance $d$
with counterions in between provides the simplest setting for studying effective interactions 
between like-charged macromolecules.
It was shown in early experiments 
\cite{Khan85,Kjellander88,Bloomfield91,Rau92,Kekicheff93},
more recently in membranes, vesicle, or bilayer systems
\cite{Komorowski18,Mukina19,Fink19,Komorowski20},
as well as in numerical simulations 
\cite{Guldbrand84,Kjellander84,Bratko86,Gronbech97}, that like-charged colloid surfaces can attract each other, under the action of Coulombic forces alone. This requires that the coupling be strong enough.
In the case of pointlike counterions, the relevant theory involves only 
one dimensionless thermodynamic parameter, namely the coupling constant $\Xi$ \cite{rque1}.
\remma{However, for many systems, one cannot ignore the finite size of the ions. These includes for example systems with bulky counter-ions like ionic liquids \cite{Gebbie13, Valmacco15}, highly charged surfaces like calcium silicate hydrates where the size of the (hydrated) ions are comparable to the average distance between neighboring surface charges \cite{Labbez06}, systems with high salt concentrations \cite{Smith16} or with dielectric discontinuities \cite{Jing15}.} It is the purpose of the present paper to go beyond the model 
of point counterions, by accounting for steric effects when these ions feature a finite size.

The weak-coupling (small $\Xi$) limit \remma{at low salt concentrations} is well described by the Poisson-Boltzmann 
mean-field theory \cite{Andelman06}. 
Within a field-theoretic representation of the Coulomb fluid grand-canonical 
partition function \cite{Edwards62}, the Poisson-Boltzmann theory is the leading term in 
a systematic loop-expansion \cite{Attard88}. 
To describe the opposite strong-coupling (SC) limit, 
a virial (fugacity) expansion of the grand-canonical partition function 
in inverse powers of the coupling constant $\Xi$ was proposed 
\cite{Moreira00,Netz01,Moreira02,Kanduc07}.
In the case of a single charged surface and to leading virial SC order,
each particle moves independently of other particles in the direction 
perpendicular to the confining surfaces,
which was verified by Monte Carlo (MC) numerical simulations 
\cite{Moreira00}.      
The first SC correction to the particle density profile \cite{Netz01,Moreira02}
has the right functional form in space, but the wrong dependence of 
the prefactor on the small parameter $1/\Xi$. 
As concerns the geometry of two parallel equivalently-charged walls, 
the analytical results for the pressure are accessible only for very small
distances $d$. 

Other theoretical attempts to construct a SC theory were based on 
the ground-state Wigner crystals created by counterions
\cite{Rouzina96,Shklovskii99,Perel99}. 
In the absence of dielectric wall images, according to Earnshaw's theorem 
\cite{Earnshaw1842} the counterions stick on the wall surfaces at infinite coupling (\emph{e.g.}~temperature goes to zero).
For the one-wall geometry, they form a two-dimensional hexagonal (equilateral
triangular) Wigner crystal. 
In the case of two parallel walls, five distinct (staggered) Wigner bilayers 
I-V were detected as the distance between the walls increases from zero 
to infinity 
\cite{Falko94,Esfarjani95,Goldoni96,Schweigert99,Weis01,Messina03,Lobaskin07}. 
Some elusive features of critical properties 
were revisited in Ref. \cite{Samaj12a} 
by using an analytic approach based on an expansion of the energy of 
the five structures in generalized Misra functions \cite{Misra}.
A SC theory based on the harmonic approximation for particle deviations from 
their ground-state Wigner positions was proposed 
in Ref. \cite{Samaj11}.
The leading order for the density profile and the pressure turns out to be 
identical to the virial single-particle theory. 
For the one-wall geometry, the first correction to the particle density profile 
has the correct functional form in space and the prefactor, 
proportional to $1/\sqrt{\Xi}$, is in good agreement with MC data.
As concerns the two-wall geometry, the harmonic analysis in 
Ref. \cite{Samaj11} is also restricted to very small distances
\cite{Varenna}.

Taking the harmonic approximation in full \cite{Samaj18}, i.e. with no restriction to small distances, leads to an effective 
one-body potential acting on particles at each of the two walls which
interpolates correctly between zero for distances $d$ much smaller than 
the Wigner lattice spacing and the locally linear potential of separated 
charged walls at asymptotically large inter-wall distances $d\to\infty$. 
This is why the profile of the particle density and the pressure exerted 
on the walls are described well also for intermediate distances between 
the walls comparable with the lattice spacing of the Wigner crystal.
The technique was first applied to asymptotically large values of the coupling 
constant when the system is in a crystal phase.
The aspect ratio of the Wigner bilayer structure (around which 
the harmonic expansion is made) was taken as a free parameter, determined by minimizing the total free energy. 
For treating the fluid phase present at smaller and more realistic values of the coupling constant $\Xi$, the Wigner bilayer structure was substituted by 
a correlation hole, i.e. a depletion region around each particle due to 
the strong Coulomb repulsion 
\cite{Rouzina96,Chen06,Nordholm84,Santangelo06,Hatlo10,Bakhshandeh11,Palaia18}. It is noteworthy that the details of the structure, crystalline versus strongly modulated liquid, only affect fine details of the effective force, and are rather immaterial.

The aim of this paper is to extend the strong-coupling two-walls analysis
of Ref. \cite{Samaj18} beyond point-like counterions, 
treating this species as charged hard spheres of
diameter $d_{\rm hc}$ which are impenetrable to other particles 
as well as to hard walls (primitive model).
Theoretical treatment of pure hard-core systems  is usually based on a potential
of mean force, the so-called depletion potential \cite{Lek11}. 
The phase diagram of hard spheres (without any charges) between parallel plates was calculated 
by using MC simulations in a wide range of particle densities and for plate 
separations ranging from one to two hard-core diameters in 
Ref. \cite{Schmidt97}.
Besides the standard fluid phase, pure hard spheres freeze into closed-packed 
versions of the crystal bilayer structures which take place 
also in the pure Coulomb problem, namely one triangular layer (phase I), 
the linear buckling structure (phase II), two square layers (phase III), 
the rhombic structure (phase IV) and two triangular layers (phase V).   
The primitive model, including both Coulomb and hard-core interactions,
was studied mainly numerically by using MC simulations, within an isolated electric double-layer
\cite{Borukhov97,Patra02,Kilic07} as well as two-wall geometry 
\cite{Valleau91,Kjellander92,Zelko10}, 
for weak and intermediate values of the coupling constant, 
up to $\Xi\simeq 100$.
Here, we shall assume that the coupling constant $\Xi$ is large,
so that the Coulomb interactions dominate in creating the ground state.
For small and intermediate inter-plate distances, the particles are supposed 
to form basically the Wigner bilayer structure of type I, II or III, their centers 
being at distance $d_{\rm hc}/2$ from either plate 1 or 2. 
We shall look, both analytically and numerically, for steric hard-sphere 
effects on this structure.

The paper is organized as follows.
A short recapitulation of the SC theory for pointlike particles,
as developed in Ref. \cite{Samaj18}, is presented in Sec. \ref{Sec2}.
Subsec. \ref{Subsec2a} reviews the relevant ground-state Wigner bilayers
while Subsec. \ref{Subsec2b} deals with the leading SC 
description of thermodynamics and the density profiles.
Sec. \ref{Sec3} generalizes the SC theory to account for ionic hard core. 
Subsec. \ref{Subsec3a} brings a list of steric restrictions on 
the parameters of the Wigner bilayers.
Subsec. \ref{Subsec3b} deals with SC thermodynamics of hard spheres.
The comparison of the theory with our Monte-Carlo numerical results 
is given in Sec. \ref{Sec4} and 
Sec. \ref{Sec5} is for the Conclusion.

\renewcommand{\theequation}{2.\arabic{equation}}
\setcounter{equation}{0}

\section{Pointlike particles} \label{Sec2}
We start with the definition of the model where positions are denoted by ${\bf r}=(x,y,z)$. 
There are two parallel plates $\Sigma_1$ at $z=0$ and $\Sigma_2$ at $z=d$
with infinite surfaces 
$\vert \Sigma_1 \vert = \vert \Sigma_2 \vert = S\to\infty$ 
along the $(x,y)$ plane.
The plates are charged symmetrically by the uniform surface charge density 
$e \sigma$ with $e$ being the elementary charge and $\sigma>0$.
For this case the resulting electric field vanishes between the plates. 

There are $N$ mobile particles between the plates, each with a  
charge $-qe$, coined as ``counterions''. The valency $q$ takes integer values
(e.g. $q=1$ for Na$^+$ ions, $q=2$ for Mg$^{2+}$ etc.) while $e$ is the electron charge.
At this stage we consider classical (i.e. non-quantum) particles 
to be pointlike.
The electro-neutrality of the system is ensured by the equality 
\begin{equation}
N q = 2\sigma S .
\end{equation}

The dimensionless distance between the plates is defined as
\begin{equation}
\eta = d \sqrt{\sigma/q} .
\end{equation}
Technically speaking, it is convenient to have a rescaled measure of distance that is temperature independent. This avoids singularities when studying the ground state, that is met under infinite coupling, see below.
The particles are immersed in a solution of dielectric constant $\epsilon$,
the same as that of the walls,
so that there are no image forces at work.
In Gaussian units, the Coulomb potential at distance $r$ is given by
$1/(\epsilon r)$. 
The system of charged particles and plates is in thermal equilibrium.

\subsection{Ground state} \label{Subsec2a}
\begin{figure}[tbp]
\includegraphics[width=0.37\textwidth,clip]{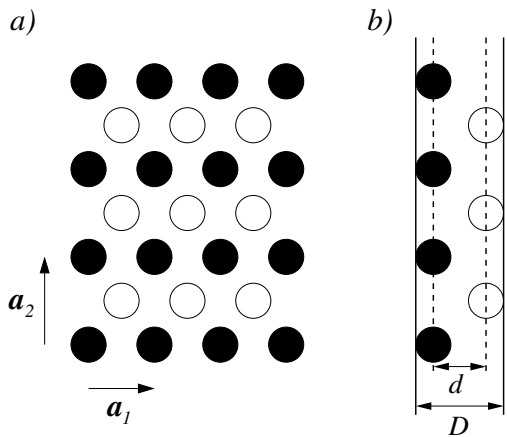}
\caption{a) Geometry for the ground-state structures I, II and III of counterions on two 
equivalently charged plates, and definition of lattice vectors
$(\bm{a}_1,\bm{a}_2)$.
Open and filled symbols correspond to particle positions on 
the opposite surfaces. The ratio $|{\bm a}_2|/|{\bm a}_1|$ defines $\Delta$. b) Side view, with definition of relevant distances. The
dimensionless distance $\eta$ between the plates is defined as $d/\sqrt{a_1 a_2}$. We have $d=D-\dhc$, where $\dhc$ is the ionic diameter and $D$ the true distance between the walls; 
$d$ turns out to be a more relevant quantity than $D$.}
\label{fig:Structures}
\end{figure}

In the ground state, corresponding to infinite coupling ($\Xi\to\infty$), our interacting point charges in a slab domain stick to the domain's boundary 
\cite{Earnshaw1842}. 
In the case of symmetrically charged plates, $N/2$ particles stick on plate 
$\Sigma_1$ and the remaining $N/2$ particles stick on plate $\Sigma_2$.
As $\eta$ goes from 0 to $\infty$, numerical simulations
\cite{Falko94,Esfarjani95,Goldoni96,Schweigert99,Weis01,Messina03,Lobaskin07}
indicate five distinct bilayer Wigner structures. 
For small and intermediate values of $\eta$ studied in this paper,
the staggered rectangular structures I, II and III are relevant.
As is shown in Fig. \ref{fig:Structures}, a single layer consists in 
the rectangular lattice with the aspect ratio $\Delta$, defined by 
the primitive translation vectors $\bm{a}_1 = a(1,0)$ and 
$\bm{a}_2 = a(0,\Delta)$. 
The lattice spacing $a$ is determined by the electroneutrality requirement
that the total surface charge in a rectangle must compensate
the charge of just one particle, $a = \sqrt{q/(\sigma\Delta)}$.
The identical rectangular structures on the two plates are shifted with 
respect to one another by a half period $(\bm{a}_1 + \bm{a}_2)/2$.

Structure I with $\Delta=\sqrt{3}$ corresponds to a (equilateral) triangular 
lattice which appears in the monolayer limit $\eta\to 0$.
The aspect ratio is from the interval $1<\Delta<\sqrt{3}$ for soft structure 
II and $\Delta=1$ for structure III which is the staggered square bilayer.
The phase transformation I--II takes place just at $\eta=0$ 
\cite{Messina03,Samaj12a}, the phase transition between structures II and 
III appears at $\eta\sim 0.263$ and phase III provides the lowest energy 
up to $\eta\sim 0.621$.

Using techniques introduced in Ref. \cite{Samaj12a}, the energy per particle 
$e_0=E_0/N$ is expressible for all three structures I-III in terms
of the generalized Misra functions
\begin{equation} \label{znu}
z_{\nu}(x,y)=\int_0^{1/\pi} \frac{{\rm d}t}{t^{\nu}}{\rm e}^{-xt}{\rm e}^{-y/t};
\end{equation}
the ordinary Misra functions correspond to $x=0$ \cite{Misra}.
In particular, writing
\begin{equation} \label{e0}
e_0(\eta,\Delta) \,= \, q^{3/2} \, \frac{e^2 \sqrt{\sigma}}{\epsilon} 
\frac{1}{2\sqrt{\pi}} \Sigma(\eta,\Delta) ,  
\end{equation}
the function $\Sigma(\eta,\Delta)$ is expressed as an infinite series 
of the generalized Misra functions in Eq. (\ref{sigmaMisra}) 
of Appendix \ref{app:A}.
The generalized Misra functions $z_{\nu}(x,y)$ with half-integer indices 
can be written in terms of the complementary error function, 
see Eqs. (\ref{zerror}) and (\ref{znu0y}) of Appendix
\ref{app:A}.
This makes the use of symbolic calculation softwares very efficient.
In practice, the infinite series (\ref{sigmaMisra}) over $(j,k)$
indices must be truncated at some $M$.
For the well known case of the hexagonal lattice with $\eta=0$ and 
$\Delta=\sqrt{3}$, the truncation of the series at $M=1,2,3,4$ 
reproduces the Madelung constant up to $2,5,10,17$ decimal digits, 
respectively \cite{Samaj12a}.
To maintain a high accuracy of our results, we truncate all Misra series 
at $M=6$.
The calculation of one ground-state energy value takes less than one 
second of CPU time on a standard PC. 

For a given distance $\eta$, the value of the rectangular aspect ratio 
$\Delta$ is determined by the energy minimization condition
\begin{equation} \label{Delta0}
\frac{\partial}{\partial\Delta} e_0(\eta,\Delta) = 0 . 
\end{equation}
This condition sets the dependence of the aspect ratio
on the dimensionless distance between the plates in the ground-state
$\Delta_0(\eta)$, see Ref. \cite{Samaj12a}.

\subsection{Crystal phase at strong coupling} \label{Subsec2b}
The system being in thermal equilibrium at some (inverse)
temperature $\beta=1/(k_{\rm B}T)$, there are two relevant length scales.
The distance at which two elementary charges interact with thermal energy 
$k_{\rm B}T$ is the Bjerrum length
\begin{equation}
\ell_{\rm B} = \frac{\beta e^2}{\epsilon} .
\end{equation}
A charge $qe$ at distance $z$ from a wall with the surface charge density 
$e\sigma$ has the potential energy $2\pi e^2 q \sigma z/\epsilon$.
The distance at which the charge $qe$ has the potential energy equal to 
thermal energy $k_{\rm B}T$ is known as the Gouy-Chapman length
\begin{equation}
\mu = \frac{1}{2\pi\ell_{\rm B}\sigma q} .
\label{eq:GouyChapman}
\end{equation}
The coordinate $z$, which is perpendicular to the charged surfaces of 
the walls, will be often expressed in units of $\mu$, $\widetilde{z}=z/\mu$.
The dimensionless coupling parameter $\Xi$, quantifying the strength
of electrostatic correlations, is defined as the ratio of the two
relevant lengths:
\begin{equation}
\Xi \,=\, \frac{q^2\ell_{\rm B}}{\mu} \, =\,  2\pi \ell_{\rm B}^2 \,\sigma \, q^3 .
\end{equation}
The strong-coupling (SC) regime $\Xi\gg 1$ is in practice most conveniently met by increasing the valence $q$. In doing so,\textbf{\textbf{\textbf{}}} excluded volume effects become prevalent, and the point-like limit of early studies less relevant. 
Alternatively, the regime of strong coupling corresponds to either 
low temperatures (a limit that is of little practical interest in view 
of applications with water, due to the unavoidable freezing of the solvent),
or large surface charge densities. 
The lattice spacing $a$ of the Wigner structure, which is the characteristic 
length scale in the longitudinal $(x,y)$ plane, is much 
larger than $\mu$ in the SC regime as $a/\mu\propto \sqrt{\Xi}$. 
In the remainder, we take $q=1$, without loss of generality,
in order not to clutter formulas. 

For a single-layer Wigner crystal, experiments \cite{Grimes79} and 
simulations \cite{Morf79} give the estimate $\Xi\gtrsim 3\times 10^4$ 
for the coupling parameter at melting from the ordered crystal to a fluid phase.  
The coupling parameter at melting of the Wigner bilayer crystal depends on $\eta$
\cite{Goldoni96}. 
Let $\Xi$ be large enough to localize particles near their Wigner-crystal 
positions.
Within the canonical ensemble, the relevant thermodynamic quantities are 
the partition function $Z_N$ and the corresponding (dimensionless) 
free energy per particle $\beta f=\beta F/N$ which are defined, up to some 
irrelevant constants due to the interaction of surface charge densities with 
themselves and charged particles, as follows
\begin{equation} \label{partition}
Z_N = \frac{1}{N!} \int_{\Lambda} \prod_{i=1}^N \frac{{\rm d}^3r_i}{\lambda^3}\,
{\rm e}^{-\beta E(\{ {\bf r}_i\})} , \quad
\beta f = - \frac{1}{N} \ln Z_N ,
\end{equation} 
where $E(\{ {\bf r}_i\})$ is the Coulomb interaction energy of the particles
and $\lambda$ stands for the thermal de Broglie wavelength.
We recall that the electric potential induced by the symmetrically charged 
plates is constant between the plates. 
The mean particle number density at point ${\bf r}$ is defined as 
$\rho({\bf r}) = \left\langle \sum_{i=1}^N 
\delta({\bf r}-{\bf r}_i) \right\rangle$,
where $\langle \cdots \rangle$ means the statistical average over 
the canonical ensemble.
It fulfils the conservation condition
$\int \rho = N$ . 
We here study the $(x,y)$-averaged density profile $\rho(z)$, which depends only on the perpendicular $z$-coordinate,
$\rho({\bf r}) \equiv \rho(z)$, so that
\begin{equation} \label{norma}
\int_0^d {\rm d}z\, \rho(z) = \frac{N}{S} = 2\sigma . 
\end{equation}
With the rescaled particle number density 
\begin{equation} \label{respart}
\widetilde{\rho}(\widetilde{z}) \equiv 
\frac{\rho(\mu\widetilde{z})}{2\pi\ell_{\rm B}\sigma^2} ,
\end{equation}
the electro-neutrality condition (\ref{norma}) takes the form
\begin{equation} \label{normatilde}
\int_0^{\widetilde{d}} {\rm d}\widetilde{z}\, \widetilde{\rho}(\widetilde{z}) = 2 . 
\end{equation}

The strong-coupling approach to the counterion system is based on 
a harmonic expansion of the energy $E$ with respect to 
particle coordinates around their 
ground state Wigner bilayer positions \cite{Goldoni96}, where the ground state corresponds
to infinite coupling.
Numerical simulations in Ref. \cite{Samaj18} indicate that at finite although large coupling, the particles form another reference crystal of type I-III 
with the aspect-ratio parameter $\Delta$ which depends, besides 
the inter-plate distance $\eta$, also on the coupling constant $\Xi$, 
i.e. $\Delta(\Xi,\eta)$.
We have performed the full harmonic expansion of particle coordinates 
around this reference crystal and fixed $\Delta(\Xi,\eta)$ of 
the reference crystal by minimizing the free energy with respect to 
this parameter.
In this paper, we keep only the leading terms linear in $z$; it turns out that
the harmonic deviations in the crystal $(x,y)$ plane as well as quadratic
terms in the $z$-direction (proportional to $1/\sqrt{\Xi}$) have only minor
effects on the results in the SC regime.
The neglect of these terms will enable us to include the hard-core
interactions in a relatively simple way.
The total energy is thus expressed as
\begin{equation} \label{totalenergy}
E(\{ {\bf r}_i\}) = N e_0(\eta,\Delta) + \delta E , 
\end{equation}
where the energy change is given by
\begin{equation} \label{deltaE}
\beta \delta E = \kappa(\eta,\Delta) \left[ \sum_{i\in\Sigma_1} \widetilde{z}_i 
+ \sum_{i\in\Sigma_2} (\widetilde{d}-\widetilde{z}_i) \right] + \cdots . 
\end{equation}
Here, the prefactor to small deviation terms is given by
\begin{eqnarray}
\kappa(\eta,\Delta) & = & \frac{\eta}{2\pi} \sum_{i_x,i_y} 
\frac{\Delta^{3/2}}{\left[ \left( i_x-\frac{1}{2}\right)^2 
+ \Delta^2 \left( i_y-\frac{1}{2}\right)^2
+\Delta\eta^2\right]^{3/2}} \nonumber \\ & = &  
- \frac{1}{2 \pi^{3/2}} \frac{\partial}{\partial\eta} \Sigma(\eta,\Delta) + 1 .
\label{kappa}
\end{eqnarray}
The leading terms are linear in $\widetilde{z}_i$ for particles sitting in 
the ground state on plate $\Sigma_1$ and in $(\widetilde{d}-\widetilde{z}_i)$ 
for particles $i\in \Sigma_2$.
The function $\kappa$ can be viewed as an effective electric one-body field due to 
the uniform surface charges on the two plates and the particle ground-state 
layer on the opposite plate.
For $\eta\to 0$ we have $\kappa\to 0$, i.e. each particle feels the zero 
field coming from the uniform surface charges on the plates while 
the effect of the opposite particle layer with the lattice spacing $a\gg d$ 
is negligible. 
For $\eta\to\infty$ we have $\kappa\to 1$, i.e. each particle feels the 
field coming from the surface charge at its own plate while the discrete 
counterion structure on the opposite plate is smeared out and neutralized by 
the opposite surface charge density on that plate.
The function $\kappa$ thus reflects a continuous interpolation between 
the two-plate case for small $\eta$-values and the one-plate case for large 
$\eta$-values.

The partition function (\ref{partition}), with the particle interaction energy 
given by Eqs. (\ref{totalenergy}) and (\ref{deltaE}), reads as
\begin{equation}
Z_N = \frac{1}{N!} \left( \frac{\mu}{\lambda} \right)^N 
\exp\left[ -\beta N e_0 \right] Q_z ,
\end{equation}
where 
\begin{eqnarray}
Q_z(\eta,\Delta) & = & \int_0^{\widetilde{d}} \prod_{i\in\Sigma_1} {\rm d}\widetilde{z}_i\,
{\rm e}^{-\kappa \widetilde{z}_i} \int_0^{\widetilde{d}} \prod_{i\in\Sigma_2} {\rm d}\widetilde{z}_i\,
{\rm e}^{-\kappa(\widetilde{d}-\widetilde{z}_i)} \nonumber \\
& = & \left( \frac{1 - \exp(-\kappa\widetilde{d})}{\kappa} \right)^N . \label{Qz} 
\end{eqnarray} 
Neglecting irrelevant terms which do not depend on $\eta$ and $\Delta$,
the leading SC representation of the free energy per particle is given by
\begin{equation} \label{freeenergy}
\beta f(\eta,\Delta) = \frac{\sqrt{\Xi}}{2^{3/2}\pi} \Sigma(\eta,\Delta) -
\ln \left( \frac{1-{\rm e}^{-\kappa(\eta,\Delta)\widetilde{d}}}{\kappa} \right) .
\end{equation}
The dependence of the aspect ratio $\Delta$ on the coupling constant 
$\Xi$ and the plate distance $\eta$, $\Delta(\Xi,\eta)$, is fixed by
the principle of minimum free energy, i.e.,
\begin{equation} \label{varDelta}
\frac{\partial}{\partial \Delta} \beta f(\eta,\Delta) = 0 .
\end{equation}
This condition is the analogue of the infinite coupling relation (\ref{Delta0}).

The pressure can be obtained via the thermodynamic route as follows
\begin{equation}
\beta P_{\rm th} = - \frac{\partial}{\partial d} 
\left( \frac{\beta F}{S} \right)
= - 2 \sigma^{3/2} \frac{\partial (\beta f)}{\partial\eta} .
\end{equation}
The pressure, rescaled as the particle density in (\ref{respart}), 
is given by
\begin{equation} \label{Pth}
\widetilde{P}_{\rm th} \equiv \frac{\beta P_{\rm th}}{2\pi\ell_{\rm B}\sigma^2}
= - \sqrt{\frac{2}{\pi\Xi}} \frac{\partial}{\partial\eta} 
\left[ \beta f(\eta,\Delta) \right] .
\end{equation}

To assess the consistency of the result, it is appreciable to have an alternative route for computing the pressure. It is offered by the contact theorem \cite{contact},
that requires the knowledge of the contact ionic density. 
The particle density profile is derived in appendix \ref{app:B}.
The contact theorem for planar walls relates the total
contact density of particles on the wall and the pressure:
\begin{equation} \label{ct}
\widetilde{P}_{\rm c} = \widetilde{\rho}(0) - 1 
= \kappa \left( \frac{1+{\rm e}^{-\kappa\widetilde{d}}}{1-{\rm e}^{-\kappa\widetilde{d}}} 
\right) - 1 .
\end{equation}
The thermodynamic $\widetilde{P}_{\rm th}$ and contact $\widetilde{P}_{\rm c}$ pressures 
in general do not coincide in an approximate theory, although they refer to the same quantity.
Their difference reveals 
the accuracy of the approach. It should be kept in mind 
that in \eqref{ct}, the local field $\kappa$ is distance dependent.

\renewcommand{\theequation}{3.\arabic{equation}}
\setcounter{equation}{0}

\section{Hard spheres: Analytical Theory} \label{Sec3}

\subsection{Steric restrictions} \label{Subsec3a} 
After having presented the key aspects of the theory for point ions, 
we now address hard-core effects: each ion is a hard-sphere of diameter 
$d_{\rm hc}$.
The hard core is impenetrable to other particles (a model referred to as the primitive model) as
well as the wall.
We shall assume that the coupling constant $\Xi$ is very large,
so that the Coulomb interactions dominate and a \remma{simple} crystal phase \remma{(I, II, or III as in the point case with only one ion per lattice cell)} is formed,
as long as it does not lead to ionic overlap. \remma{Scanning only these simple crystal phases was further motivated by viscual inspection of the structures found by our Monte Carlo simulations}.
The counterions are supposed to be close to the Coulomb bilayer structure 
of type I-III, their centers being at distance $d_{\rm hc}/2$ from either of
plates 1 or 2 and we shall look for steric hard-sphere effects on 
this structure. 

If $D$ is the true distance between the walls, it is useful to define
the reduced distance $d$ via $D=d+d_{\rm hc}$, where $d$ is the distance available to the center of mass of hard-sphere ions; it is equal to
0 in the extreme case when particles touch by their hard-core surfaces  
simultaneously both plates, see Fig. \ref{fig:Structures}.
As above, we use the notation $\eta=d\sqrt{\sigma}$. 
It is useful to express lengths in terms of the lattice spacing $a_b$ of 
the hexagonal Wigner bilayer at $\eta=0$ (with $\Delta=\sqrt{3}$), given by
\begin{equation}
\frac{\sqrt{3}}{2} a_b^2 = \frac{1}{2\sigma} .
\end{equation}
Structure I can exist at $\eta=0$ only if $d_{\rm hc}\le a_b$.
It is therefore natural to introduce the parameter
\begin{equation}
r \,\equiv\, \frac{d_{\rm hc}}{a_b} 
\,=\, d_{\rm hc} \sqrt{\sigma} \, 3^{1/4} \,=\,\frac{3^{1/4}}{\sqrt{2\pi\Xi}} 
\widetilde{d}_{\rm hc}
\end{equation}
which compares Coulomb and steric effects in the system.
Note that $r=1$ when $d_{\rm hc} \sqrt\sigma = 3^{-1/4} \simeq 0.76$.
When $r<1$, the main expectation goes as follows. 
If $\Xi$ is sufficiently large, the ions strongly repel each other,
so that their ``in-plane'' ($xy$) motion is essentially frozen: their only possible motion takes place perpendicularly to the plates, along $z$.
It is consequently immaterial to consider point-ions, or hard-sphere ions,
as long as $r<1$. We then expect that when expressed in terms of the $d$ 
variable, the pressure curves should be independent of the ionic diameter.
This ``no-hindrance regime'' will be illustrated in section \ref{Sec4}.

When $r>1$, steric hindrance impinges on the point-like arrangement, and needs to be properly addressed.
Supposing that the counterions form basically the Coulomb bilayer structure 
of type I-III, there are strong steric hard-sphere restrictions on model
parameters which have both intra-layer and inter-layer nature. 
We start by the intra-layer analysis.
The existence of the structures I-III is limited by the condition 
$d_{\rm hc}\le a(\Delta) = 1/\sqrt{\sigma\Delta}$ which implies 
the restriction
\begin{equation} \label{Deltarestr}
\Delta \le \frac{\sqrt{3}}{r^2} .
\end{equation}
If $r\in [1,3^{1/4}]$, the formula (\ref{Deltarestr}) yields a restriction on 
the parameter $\Delta$.  
For $r>3^{1/4} \simeq 1.316$, the bilayer Wigner structures I-III cannot exist at all. 

\begin{figure}[tbp]
\includegraphics[width=0.35\textwidth,clip]{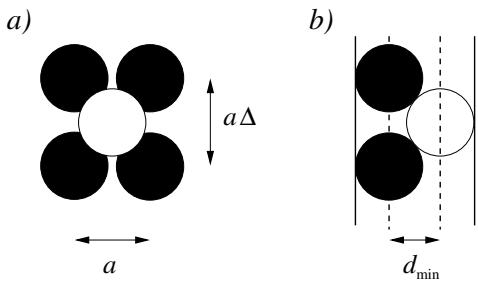}
\caption{Closest approach configuration where the distance between 
the centers of two nearest neighbor ions of different colors equals the hard core diameter $\dhc$. Given the lengths reminded in the picture, one obtains the close packing condition \eqref{eq:closepacking}.
a) In-plane view; b) Side view. 
The distance between the two dashed lines is $d_{\rm min}$;
it is the minimum value $d$ can take. For a fixed $\Delta$, the ions have here no free volume.}
\label{fig:closepacking}
\end{figure}

\begin{figure}[tbp]
\includegraphics[width=0.4\textwidth,clip]{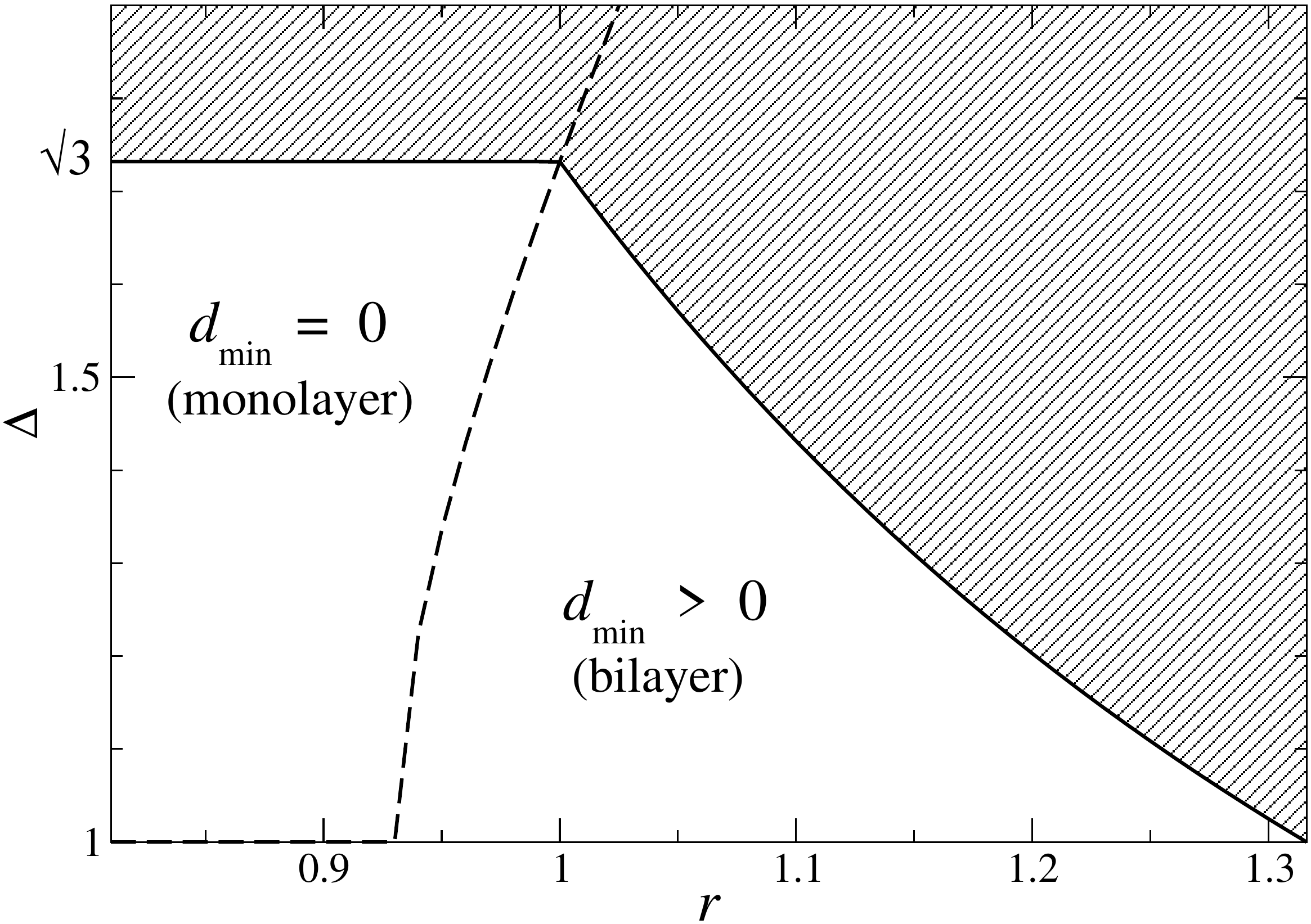}
\caption{Aspect ratio $\Delta$ as a function of reduced hard core ionic diameter $r=\dhc/a_b$. The hatched region is for the forbidden $\Delta$ values that exceed the intra-plate bound \eqref{Deltarestr}, or that exceed $\sqrt{3}$. The dashed line shows $\Delta^*$ defined in 
\eqref{eq:Deltastar}. It discriminates a region where the minimal 
distance between plates can vanish in the $d$ variable (corresponding to an inter-plate distance equal to the ionic diameter, and thus a monolayer), from another where steric effects preclude this possibility and lead to a non-vanishing minimal distance $d$, as defined in Fig. \ref{fig:Structures}. The interpretation of the $r=1$ threshold is that it corresponds to the maximum hard core size compatible with a possible compaction of the system down to $d=0$; the monolayer is then triangular (hexagonal),
with $\Delta=\sqrt{3}$. The analysis is here restricted to monolayer or bilayer upon contact, discarding situations with more than three layers, that would be formed in the
hatched region.}
\label{fig:Delta}
\end{figure}

\begin{figure}[tbp]
\includegraphics[width=0.4\textwidth,clip]{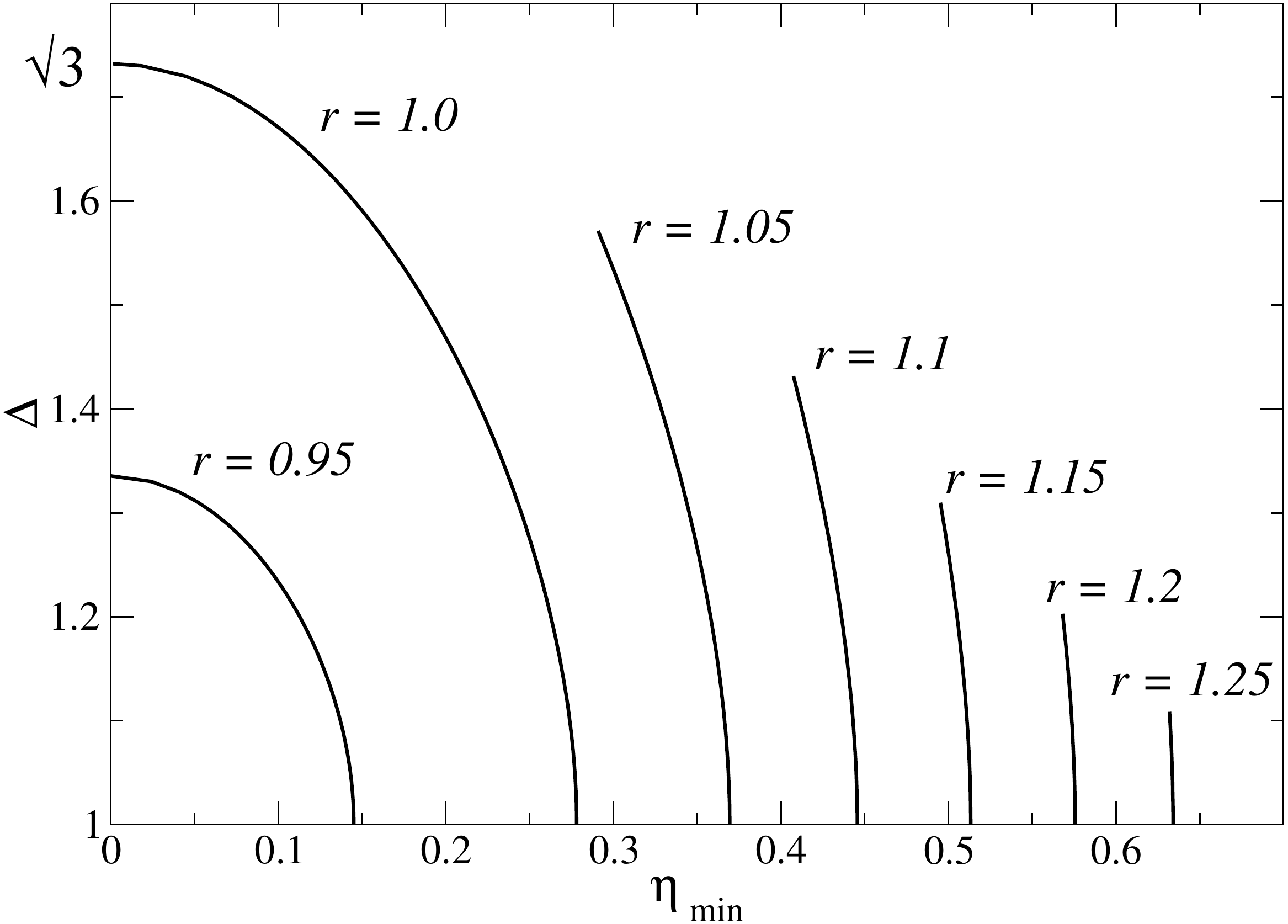}
\caption{Connection between structural aspect ratio $\Delta$ and
rescaled minimal distance $\eta_{\min}=d_{\min} \sqrt{\sigma}$ for different values of $r$, as indicated. These graphs can be viewed as vertical cuts in Fig. \ref{fig:Delta}. For $r<1$, all state points on the right hand side of the curve are accessible, see Fig. \ref{fig:etamin_Delta_comment}; for $r>1$, the additional constraint of having $\Delta$ below the highest one reported, given by Eq. \eqref{Deltarestr}, should be enforced as well.}
\label{fig:etamin_Delta}
\end{figure}

\begin{figure}[tbp]
\includegraphics[width=0.4\textwidth,clip]{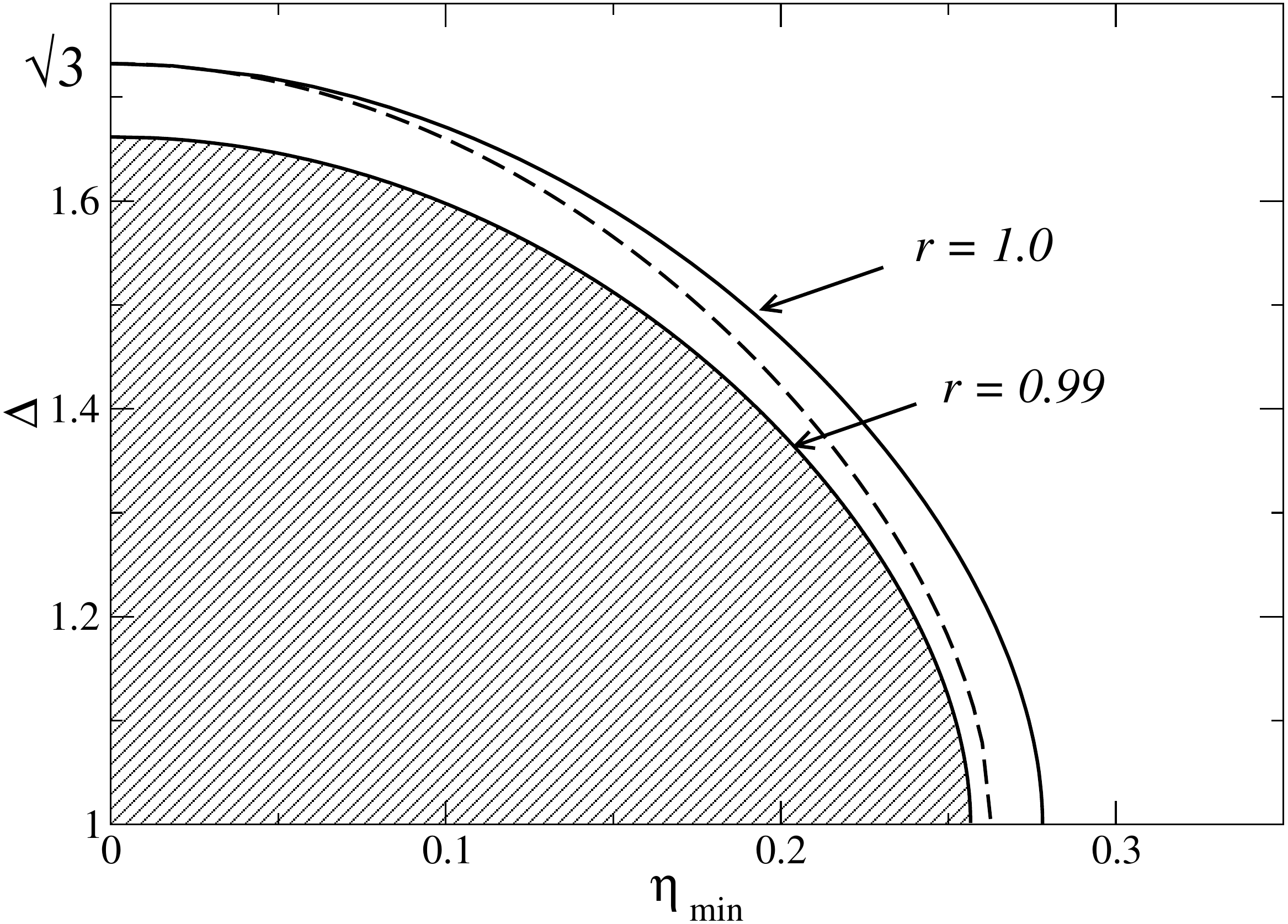}
\caption{Construction of the forbidden region (hatched) for $r=0.99$. The $\eta_{\min}$ curve for $r=1$, shown in Fig. \ref{fig:etamin_Delta}, is also reported. The dashed line corresponds to the ground state optimal
configuration when $r=0$, i.e. for point-like ions, as derived in 
\cite{Samaj12a}.}
\label{fig:etamin_Delta_comment}
\end{figure}

As concerns the inter-layer hard-core restrictions on structures I-III, 
there exists a minimal distance $d_{\min}$ at which the two layers can
approach one another.
This distance is determined as the one at which two nearest-neighbor
hard-core particles from the opposite layers touch one another (see Fig. \ref{fig:closepacking}):
\begin{equation}
d_{\rm hc}^2 = \left( \frac{\Delta a}{2} \right)^2 
+ \left( \frac{a}{2} \right)^2 + d_{\min}^2 .
\label{eq:closepacking}
\end{equation}
Equivalently,
\begin{equation} \label{etamin}
\eta_{\min}^2(\Delta;r) = \frac{r^2}{\sqrt{3}} - 
\frac{1}{4} \left( \Delta + \frac{1}{\Delta} \right) . 
\end{equation}
For a fixed $r$, the right-hand-side (rhs) of this equation is 
a monotonously decreasing function of $\Delta$ ($1<\Delta$).
If $r\le 3^{1/4}/\sqrt{2} = 0.930605\ldots$, we have $\eta_{\min} = 0$, i.e.,
there is no inter-layer restriction on structures I-III.
For $r\ge 1$ it holds that $\eta_{\min}\ge 0$ for an arbitrary value of
$\Delta\in [1,\sqrt{3}]$, i.e., there is always a hard-core restriction for 
distances between layers.
For $r\in [0.930605,1]$, there is an interval of the aspect ratios
$\Delta\in [1,\Delta^*(r)]$ with $\eta_{\min}\ge 0$ and an interval of 
$\Delta\in [\Delta^*(r),\sqrt{3}]$ with $\eta_{\min} = 0$, 
$\Delta^*(r)$ being given by
\begin{equation}
\Delta^*(r) \, =\,  \frac{2}{\sqrt{3}} r^2 + \sqrt{\frac{4}{3} r^4 - 1} .
\label{eq:Deltastar}
\end{equation}
Figure \ref{fig:Delta} summarizes the situation, showing the domain 
of validity of the different regimes in the ($\Delta,r$) plane.
Figure \ref{fig:etamin_Delta} shows how the minimum separation 
$\eta_{\min}$ and aspect ratio $\Delta$ are related, in the allowed
domain. This domain is defined differently if $r<1$ and if $r>1$,
see the caption, and also Fig. \ref{fig:etamin_Delta_comment}
which highlights the forbidden region for $r=0.99$. The reason for showing both data at $r=0.99$ and $r=1$ in Fig. \ref{fig:etamin_Delta_comment} lies in the dashed curve, that shows how the geometry of the ground state problem without hard-core
($r=0$) depends on inter-plate distance $\eta$. This curve was obtained analytically in \cite{Samaj12a}. It lies, although marginally, in the forbidden region of the $r=1$ case. Yet, it lies in the acceptable region with $r=0.99$. This allows to state that 
starting from the optimal ground state configuration of point charges, and gradually increasing the radius of hard sphere ions,
steric effects will not alter the point-like configuration for $r<0.99$. They start to do so for $r$ slightly above 0.99.

For the limiting case $\eta_{\min}(\Delta,r)=0$, the two 
plates are allowed to touch one another ($d=0$).
The equality $\eta_{\min}(\sqrt{3},r)=0$ is satisfied for $r=1$ which
is the threshold beyond which $\eta_{\min}$ is positive.
As soon as $\eta_{\min}>0$, the pressure is infinite for all 
inter-plate distances $\eta < \eta_{\min}$, since the hard spheres cannot
be packed in such a small space.

The crystal state of the counterion system now depends not only on 
the coupling constant $\Xi$ but also on the $r$-parameter whose large value
can decrease substantially the coupling constant $\Xi$ at which the 
crystal-fluid phase transition occurs.
In the crystal phase, we can treat the hard-core system basically in the same
way as the pointlike one in Sec. \ref{Sec2}, to obtain the effective 
(dimensionless) potential $-\kappa(\eta,\Delta) \widetilde{z}$ acting
on particles at plate 1 and the symmetrically reflected one with respect to
the slab center at $z=d/2$, $-\kappa(\eta,\Delta)(\widetilde{d}-\widetilde{z})$, 
acting on particles at plate 2.
Because of strong Coulomb repulsions in the $(x,y)$ plane, particles move 
freely along the lines in the perpendicular $z$-direction defined basically by 
the ground-state structures I-III.
Due to interlayer steric effects, the particles at plate 1 move in a reduced 
interval $\widetilde{z}\in [0,\widetilde{d}-\widetilde{d}_{\min}]$ while those at plate 2 
in the interval $\widetilde{z}\in [\widetilde{d}_{\min},\widetilde{d}]$. 
In what follows, we shall use the following 
combination of variables:
\begin{equation}
h(\eta,\Delta;r)  =  \kappa(\eta,\Delta) \left[ \widetilde{d}-\widetilde{d}_{\min}(\Delta;r)\right]
,
\end{equation}
where $\widetilde d = \sqrt{2\pi\Xi} \, \eta$, with a similar relation between $\widetilde d_{\min}$ and $\eta_{\min}$.

Finally, previous works \cite{Samaj11,Samaj18} have shown that even when the ionic system is not coupled enough to be in its crystal phase but rather exhibits a strongly modulated liquid structure, the large-$\Xi$ calculations are nevertheless relevant as an approximate approach. The main reason is that both
structures, liquid and solid, do exhibit the common feature of a correlation hole around each ion \cite{Moreira02,Palaia18}. We thus here develop a 
theory that is grounded in the large-$\Xi$ regime, the relevance of which at moderate couplings has to be assessed by a direct comparison to numerical simulations.

\subsection{Thermodynamics} \label{Subsec3b} 

\begin{figure}[tbp]
\includegraphics[width=0.49\textwidth,clip]{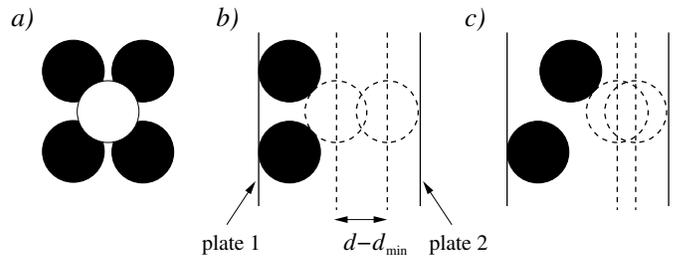}
\caption{Situation where  $d > d_{\rm min}$. Panel a) sketches the plaquette of four black ions for the tagged central (white) ion on plate 2.
The side view b) shows the accessible slab for the tagged ion, of thickness $d-d_{\rm min}$, between the vertical dashed lines.
Panel c) is for a case where one of the 4 plaquette ions (the ``upper ion'') has moved away from its ground state position, which diminishes the available space 
for the tagged ion, again materialized by the slab between the two dashed lines. It is assumed here that since the white ion only moves perpendicularly to the plate, the ``bottom'' black ion does not contribute to the available space.}
\label{fig:plaquette}
\end{figure}

To account for steric effects on the Coulomb free energy in the leading SC 
order (\ref{freeenergy}), let us take one of the particles at plate $\Sigma_2$
as the reference ion. 
It has just four nearest neighbors at the corners of one rectangular 
plaquette of the Wigner crystal at plate $\Sigma_1$; we denote these particles 
by 1,2,3,4 and their perpendicular positions respectively by
$\widetilde{z}_1$, $\widetilde{z}_2$, $\widetilde{z}_3$, $\widetilde{z}_4$.
Because the particles are supposed to move along the lines determined by Wigner
layers in the perpendicular $z$-direction, from among four positions only 
the maximal one $\max(\widetilde{z}_1,\widetilde{z}_2,\widetilde{z}_3,\widetilde{z}_4)$ 
is relevant.
The original interval $[\widetilde{d}_{\min},\widetilde{d}]$ accessible to the
reference particle is thus reduced to $[\widetilde{d}_{\min}+\max(\widetilde{z}_1,
\widetilde{z}_2,\widetilde{z}_3,\widetilde{z}_4),\widetilde{d}]$,
see Fig. \ref{fig:plaquette}.
The contribution of the reference particle on plate 2 to the partition 
function can be integrated out as follows
\begin{eqnarray}
\int_{\widetilde{d}_{\min}+\max(\widetilde{z}_1,\widetilde{z}_2,\widetilde{z}_3,\widetilde{z}_4)}^{\widetilde{d}}
{\rm d}\widetilde{z}\, {\rm e}^{-\kappa(\widetilde{d}-\widetilde{z})} 
\phantom{aaaaa} \nonumber \\ = \frac{1}{\kappa} \left[ 
1 - {\rm e}^{\kappa\max(\widetilde{z}_1,\widetilde{z}_2,\widetilde{z}_3,\widetilde{z}_4)-h} \right] ,
\end{eqnarray}
where $h=\kappa(\widetilde d-\widetilde d_{\rm min})$, is a rescaled measure of available space.
Performing the above procedure independently for every of $N/2$ particles 
on plate 2, the $N$-particle partition function reduces to the one of 
$N/2$ particles at plate 1,
\begin{eqnarray}
Q_z & = & \frac{1}{\kappa^{N/2}}
\int_0^h \prod_{i\in \Sigma_1} {\rm d}\widetilde{z}_i\, {\rm e}^{-\kappa\widetilde{z}_i}
\nonumber \\ & & \times
\prod_{{\rm plaq}(i)} \left[1 - 
{\rm e}^{\kappa\max(\widetilde{z}_{i1},\widetilde{z}_{i2},\widetilde{z}_{i3},\widetilde{z}_{i4})-h} \right] ,  
\end{eqnarray}
where the product is over all $i=1,2,\ldots,N/2$ plaquettes of the Wigner 
rectangular lattice at plate 1, the coordinates of particles localized 
at the four corners of plaquette $i$ being denoted as 
$\widetilde{z}_{i1}$, $\widetilde{z}_{i2}$, $\widetilde{z}_{i3}$ and $\widetilde{z}_{i4}$. 
We see that the elimination of one half of particles implies plaquette
four-particle interactions among the remaining half of particles. 
Finally, making the substitution $s_i=\kappa \widetilde{z}_i$ one ends up with
\begin{eqnarray} 
Q_z & = & \frac{1}{\kappa^N} \int_0^h \prod_{i=1}^{N/2} {\rm d}s_i\, {\rm e}^{-s_i}
\nonumber \\ & & \times
\prod_{{\rm plaq}(i)} \left[1 - {\rm e}^{\max(s_{i1},s_{i2},s_{i3},s_{i4})-h} \right] .  
\label{pf}
\end{eqnarray}

Through the original variables $\{s_i\}$, the plaquettes are
coupled, which makes the statistical mechanics problem at hand untractable.
We shall treat the partition function $Q_z$ approximatively by using 
the Gibbs-Bogoliubov inequality \cite{Feynman98}:
\begin{equation}
- \ln \left( {\rm Tr}\, {\rm e}^{-H} \right) \le
{\rm Tr}\left( p_0 \ln p_0 \right) + {\rm Tr}\left( p_0 H \right) ,
\end{equation} 
where $p_0$ is any normalized probability distribution, 
${\rm Tr}\, p_0 = 1$.
Comparing this formula with the studied case (\ref{pf}), we identify
the (dimensionless) Hamiltonian
\begin{equation} \label{dimHam}
H \equiv \sum_{i=1}^{N/2} s_i - \sum_{{\rm plaq}(i)} \ln \left[
1 - {\rm e}^{\max(s_{i1},s_{i2},s_{i3},s_{i4})-h} \right]
\end{equation}
and
\begin{equation}
{\rm Tr} \equiv \int_0^h \prod_{i=1}^{N/2} {\rm d}s_i .
\end{equation}
Let us choose
\begin{equation} 
p_0(s_1,s_2,\ldots,s_{N/2}) = 
\left( \frac{\alpha}{1-{\rm e}^{-\alpha h}} \right)^{N/2}
\prod_{i=1}^{N/2} {\rm e}^{-\alpha s_i} ,
\label{eq:p0ansatz}
\end{equation}
where $\alpha$ is a free (real) parameter.
The reason for this choice is dictated by the observation 
of ionic density profiles, see below, that appear essentially exponential. In other words, $\alpha$ plays the role of a multiplying factor to the local effective electric field, dressed by steric effects. In absence of hard-core interactions, one would have 
precisely $\alpha=1$, and the treatment of section \ref{Sec2} would apply. The fact that $\alpha \neq 1$ (with an effective field $\alpha \kappa$) will be a direct signature of hard-core interactions.
Our choice of trial probability $p_0$ decouples the plaquette,
as mean-field treatments do. 
Since
\begin{widetext}
\begin{equation}
{\rm Tr}\left( p_0 \ln p_o \right) = \frac{N}{2} \left[ \ln\alpha
- \ln\left( 1 - {\rm e}^{-\alpha h} \right) -
\frac{1-(1+\alpha h) {\rm e}^{-\alpha h}}{1-{\rm e}^{-\alpha h}} \right]
\end{equation}
and
\begin{equation}
{\rm Tr}\left( p_0 H \right) = \frac{N}{2\alpha}
\frac{1-(1+\alpha h) {\rm e}^{-\alpha h}}{1-{\rm e}^{-\alpha h}}
- 2 N \alpha\, \frac{\int_0^h {\rm d}s\, {\rm e}^{-\alpha s} 
\left( 1 - {\rm e}^{-\alpha s} \right)^3 \ln\left( 1 - {\rm e}^{s-h}\right)}{
\left( 1-{\rm e}^{-\alpha h}\right)^4} ,
\end{equation}
we obtain that
\begin{eqnarray}
- \frac{1}{N} \ln Q_z & \le & \ln \kappa + \frac{1}{2} \left[ \ln\alpha
- \ln\left( 1 - {\rm e}^{-\alpha h} \right) \right] +
\frac{1}{2}\left( \frac{1}{\alpha} - 1 \right)
\frac{1-(1+\alpha h) {\rm e}^{-\alpha h}}{1-{\rm e}^{-\alpha h}} \nonumber \\
& & - 2 \alpha\, \frac{\int_0^h {\rm d}s\, {\rm e}^{-\alpha s} 
\left( 1 - {\rm e}^{-\alpha s} \right)^3 \ln\left( 1 - {\rm e}^{s-h}\right)}{
\left( 1-{\rm e}^{-\alpha h}\right)^4} .
\end{eqnarray}
Consequently, the free energy of hard spheres with the Coulomb interaction
satisfies the inequality
\begin{eqnarray}
\beta f(\eta,\Delta;r) & \le & \frac{\sqrt{\Xi}}{2^{3/2}\pi} 
\Sigma(\eta,\Delta) + \ln \kappa(\eta,\Delta) + \frac{1}{2} \left[ \ln\alpha
- \ln\left( 1 - {\rm e}^{-\alpha h} \right) \right] +
\frac{1}{2}\left( \frac{1}{\alpha} - 1 \right)
\frac{1-(1+\alpha h) {\rm e}^{-\alpha h}}{1-{\rm e}^{-\alpha h}} \nonumber \\
& & - 2 \alpha\, \frac{\int_0^h {\rm d}s\, {\rm e}^{-\alpha s} 
\left( 1 - {\rm e}^{-\alpha s} \right)^3 \ln\left( 1 - {\rm e}^{s-h}\right)}{
\left( 1-{\rm e}^{-\alpha h}\right)^4} .
\end{eqnarray}
\end{widetext}
The free parameter $\alpha$ is chosen to minimize the upper bound for
the free energy, i.e., the rhs of this equation. 
In all the cases studied, the obtained $\alpha$ is in the interval $[1,\infty]$. 
The parameter $\alpha$ thus increases the slope of the decay of the particle 
density from the wall surface; since the density profile is normalized this
automatically means the increase of the particle density at the wall
as the consequence of the hard-core repulsion from particles close to 
the opposite wall. This can be thought of as a generalized depletion effect,
where ions are pushed to their nominal plate by hard core, that adds 
to the Coulomb repulsion already at work for point particles.
As before, the aspect ratio of the rectangular lattice $\Delta$ is 
also the minimizer of the free energy, respecting the hard-core 
restriction (\ref{Deltarestr}).
The (rescaled) thermodynamic pressure is given by formula (\ref{Pth}).

We turn to the density profile. Its part due to the particles in the vicinity of plate 1, 
$\rho_1(z)$, is obtained in analogy with the case of pointlike particles by 
introducing the generating Boltzmann factor $w({\bf r})$,
see Appendix \ref{app:B}.
This means that the dimensionless Hamiltonian (\ref{dimHam}) has to be
substituted in the original expression for the whole partition function
as follows $H\to H-\sum_{i=1}^{N/2} \ln w({\bf r}_i)$.
Within the Gibbs-Bogoliubov formalism, $\ln w({\bf r}_i)$ appears in
the evaluation of ${\rm Tr}(p_0 H)$ and therefore this is just the trial
distribution (\ref{eq:p0ansatz}) which determines the particle density
\begin{equation}
\widetilde{\rho}_1(\widetilde{z}) = 
\frac{\alpha\kappa}{1-{\rm e}^{-\alpha\kappa(\widetilde{d}-\widetilde{d}_{\min})}}
{\rm e}^{-\alpha\kappa\widetilde{z}} \theta(\widetilde{d}-\widetilde{d}_{\min}-\widetilde{z}) ,
\label{eq:alpha_rho1}
\end{equation}
where $\theta$ denotes the Heaviside step function.
Given the factorized form taken in Eq. \eqref{eq:p0ansatz}, this 
result does not come as a surprise.
Although particles at plate 2 have been integrated out within our approach, 
their contribution to the density profile is determined by its reflection 
$\widetilde{z}\to \widetilde{d}-\widetilde{z}$ symmetry as follows
\begin{equation}
\widetilde{\rho}_2(\widetilde{z}) = 
\frac{\alpha\kappa}{1-{\rm e}^{-\alpha\kappa(\widetilde{d}-\widetilde{d}_{\min})}}
{\rm e}^{-\alpha\kappa(\widetilde{d}-\widetilde{z}} 
\theta(\widetilde{z}-\widetilde{d}_{\min}) .
\label{eq:alpha_rho}
\end{equation}
The total density of particles is given by
\begin{equation}
\widetilde{\rho}(\widetilde{z}) = \widetilde{\rho}_1(\widetilde{z}) + 
\widetilde{\rho}_2(\widetilde{z}) .
\end{equation} 
In the analogous formula for pointlike particles (\ref{rho0}), the Coulombic
effects are expressed through the function $\kappa\in [0,1]$ which is coupled to
$\widetilde{z}$ and $\widetilde{d}-\widetilde{z}$ in exponentials.
Now there is an additional multiplication parameter $\alpha\in [1,\infty]$ 
which reflects the ``squeezing'' effect of the ionic hard core.

The contact version of the pressure follows from the contact theorem (\ref{ct}).
We have to distinguish between two cases.
If $\widetilde{d}_{\min}\le 0$, the particles from plate 2 can touch plate 1
and therefore 
\begin{eqnarray}
\widetilde{P}_c & = & \widetilde{\rho}_1(0) + \widetilde{\rho}_2(0) -1 \nonumber \\
& =  & \alpha \kappa \left( \frac{1+{\rm e}^{-\alpha\kappa\widetilde{d}}}{
1-{\rm e}^{-\alpha\kappa(\widetilde{d}-\widetilde{d}_{\min})}} \right) - 1 .
\end{eqnarray}
If $\widetilde{d}_{\min}>0$, the particles from plate 2 cannot touch plate 1
and therefore
\begin{equation}
\widetilde{P}_c = \widetilde{\rho}_1(0) -1 =  
\frac{\alpha\kappa}{1-{\rm e}^{-\alpha\kappa(\widetilde{d}-\widetilde{d}_{\min})}}-1 . 
\end{equation}

\renewcommand{\theequation}{4.\arabic{equation}}
\setcounter{equation}{0}

\section{Hard spheres: Numerical Results} 
\label{Sec4}

\renewcommand{\theequation}{5.\arabic{equation}}
\setcounter{equation}{0}

\subsection{Monte-Carlo Simulations}
\label{ssec:MC}

\begin{figure*}[t!]
\includegraphics[width=0.325\textwidth,clip]{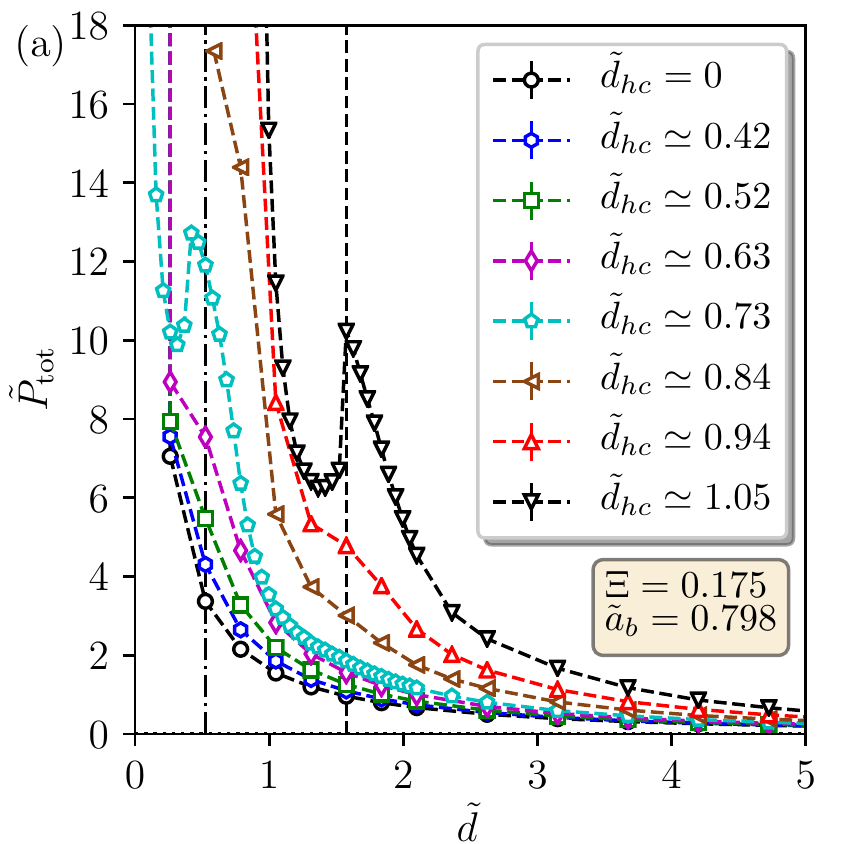}
\includegraphics[width=0.325\textwidth,clip]{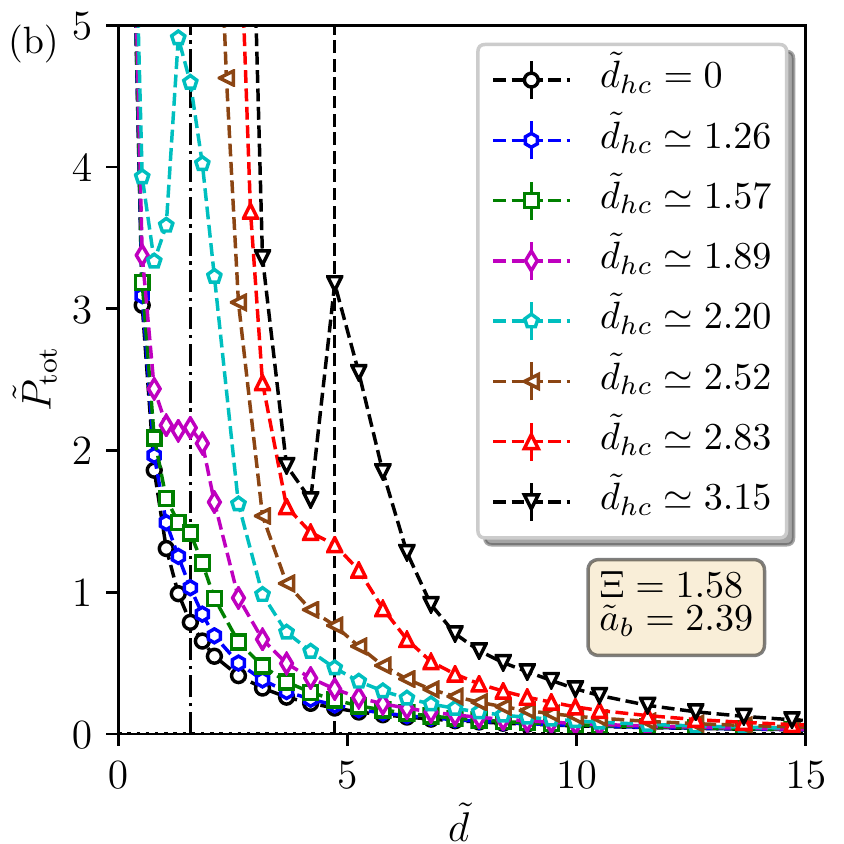}
\includegraphics[width=0.325\textwidth,clip]{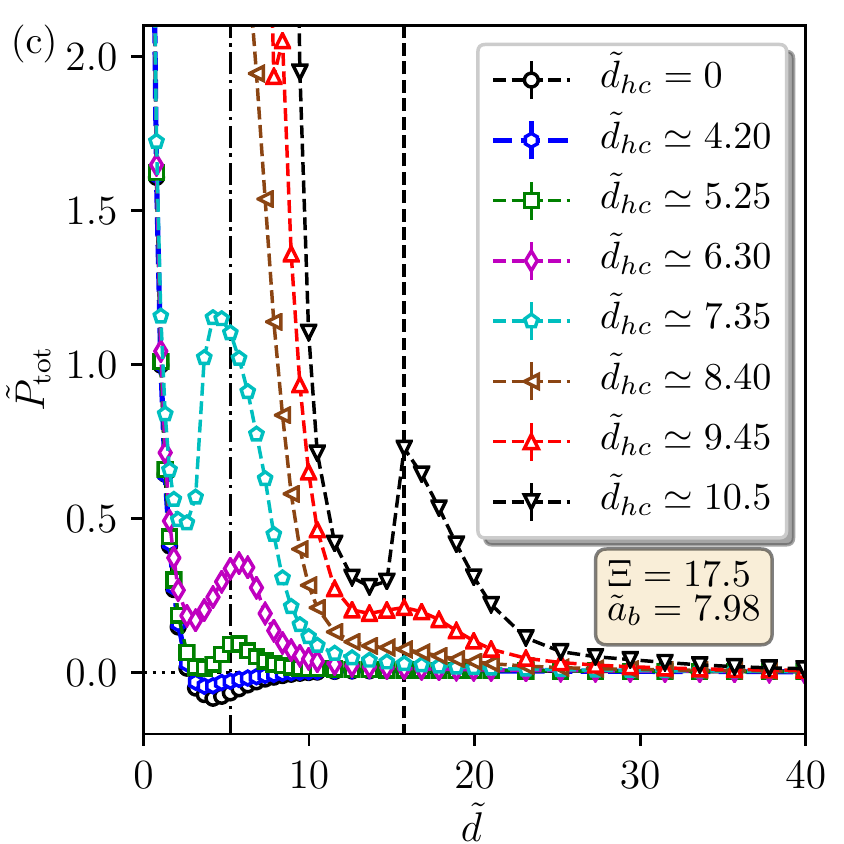}
\includegraphics[width=0.325\textwidth,clip]{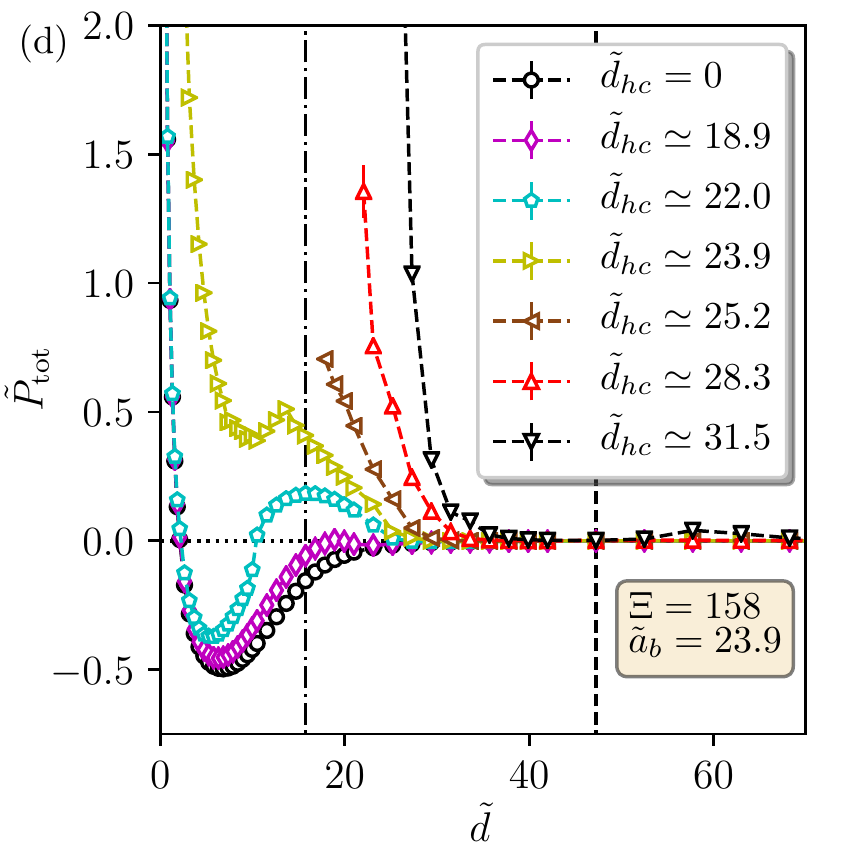}
\includegraphics[width=0.325\textwidth,clip]{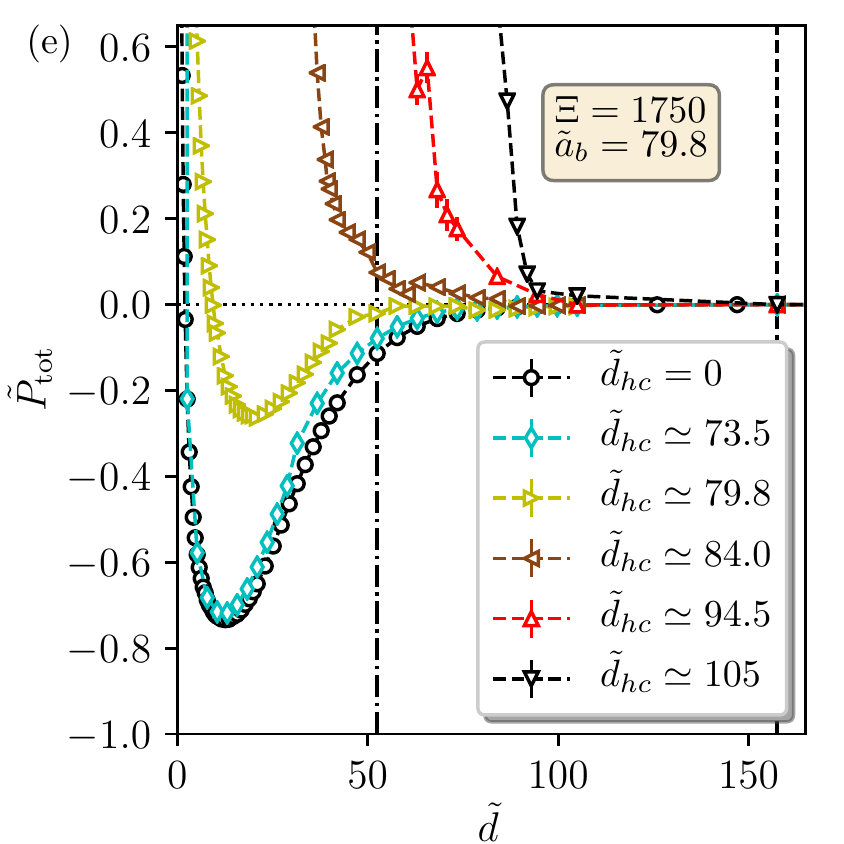}
\includegraphics[width=0.325\textwidth,clip]{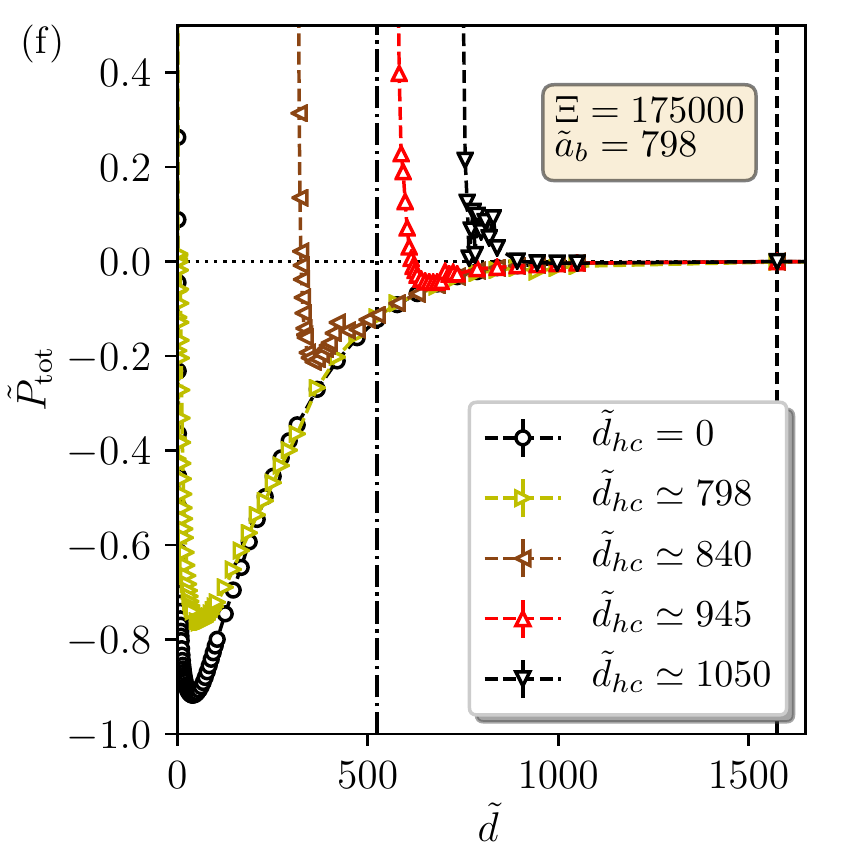}
\caption{Monte Carlo results for the equation of states at various hard core radii and coupling parameters: (a) $\Xi=0.175$ (b) $\Xi=1.58$, (c) $\Xi=17.5$, (d) $\Xi=158$, (e) $\Xi=1750$, and (f) $\Xi=175000$.
The different colors are for different values of $d_{\rm hc}\sqrt{\sigma} = r/3^{1/4}$. The values of $\widetilde{d}_{\rm hc}$ are given as well, together with $\widetilde a_b$ from which $r=\widetilde{d}_{\rm hc}/\widetilde a_b$ follows.
Black vertical lines show $\eta=0.5$ (dashed-dotted) and $\eta=1.5$ (dashed). 
In all panels, the color code is consistent: $\dhc \sqrt{\sigma}$= 0 (black circles), 0.4 (blue hexagons), 0.5 (green squares), 0.6 (maroon diamonds), 0.7 (cyan pentagons), 0.76 ($r=1$, yellow triangles pointing right), 0.8 (brown triangles pointing left), 0.9 (red triangles pointing up), and 1.0 (black triangles pointing down, for which $r=3^{1/4}$).
In each panel, the value of $a_b$ is indicated. It corresponds to 
the maximal hard core size compatible with $d=0$, or in other words, a distance $D=\dhc$ between the plates. }
\label{fig:Pressure_MC}
\end{figure*}

To put to the test the analytic theory, we run Metropolis Monte Carlo simulations of the system composed of two symmetrically charged surfaces with counterions in-between, at various coupling parameters, separations and hard core radii. For each simulation, we use 512 spherical counterions, which all have their charge located in the center of their hard core. The planar surfaces are modeled as uniformly charged structureless hard walls. Long-ranged electrostatic interactions are handled by three-dimensional Ewald summation techniques with corrections for quasi-2-dimensionality, by adding vacuum slabs on each side of the charged walls (as described elsewhere \cite{A,B,Samaj18}). We verified that our vacuum slabs were large enough in order not to  influence our results (i.e., pressures and profiles), typically larger than a couple of thousands of Gouy–Chapman lengths defined in \eqref{eq:GouyChapman}. Besides standard particle trial displacements, we also utilize floppy-box moves at constant box volume at which counterions are displaced conformally: either by shear or coupled biaxial compression-decompressions (where we compress one axis and decompress the other) both in the plane parallel to the surfaces. 
Trial move parameters were set such to have an acceptance ratio between roughly 20 and 50\% for each case.
Pressures and profiles at a fixed separation, a given counterion hard core radius, and a given coupling parameter are estimated by first equilibrating for $10^4$ Monte Carlo cycles and then sampling over $10^5$ subsequent cycles, where a cycle consists of either 512 trial counterion displacements or a trial floppy box move (a fifth of the total cycles). Pressures are evaluated both at the walls (contact theorem) or over the mid-plane by sampling the concentration (entropic contribution), ion-ion correlation (electrostatic energy), and hard-core repulsion (impulse) over the mid-plane \cite{Guldbrand84}. Both measures give the same results within statistical errors. The mid-plane evaluation is usually less noisy and hence all simulation results are reported using this measure. We apply block averaging of ten blocks to estimate the precision in pressures. 
Starting configurations for our simulations are counterion bilayers of structure I if $r<1$ otherwise structure II with $\Delta$ equal to the upper bound of Eq. (3.3), compatible with the minimum separation.

Figure \ref{fig:Pressure_MC} shows the numerical results of the equation of state for six different coupling parameters. The two first and lowest ones, $\Xi=0.175$ and $\Xi=1.58$, yield similar pressure curves.
Note that the point-like limit provides a universal equation of state, independent of $\Xi$ provided it is not too large (less than 2).
This point-like limit is here in excellent agreement with Poisson-Boltzmann theory results (not shown). Beyond point ions, steric effects result in very similar pressure curves at the two lowest $\Xi$ studied; these effects are responsible for the relevance of the $a$ parameter (or equivalently $a_b$), for scaling out results.
These two equations of states are repulsive, irrespective of the hard core radius and separation, with pressure curves increasingly repulsive when increasing hard core radii at a given separation. These low-$\Xi$ results serve as a reference to our strong-coupling analysis, illuminating the importance of increasing electrostatic coupling. Two peaks appear at these low coupling parameters, one at $\eta \simeq 0.4$ when $\dhc\sqrt{\sigma}=0.7$ (cyan symbols) and the other at $\eta \simeq 1.5$ when $\dhc\sqrt{\sigma}=1$.
They are fingerprints of the pure hard core system, in this parameter range barely affected by the electric charges. For instance, the 
change of behaviour for $\eta\simeq 0.4$ and $\dhc\sqrt{\sigma}=0.7$ is consistent with the confined hard-sphere phase diagram reported by Schmidt and L\"owen \cite{Schmidt97}.
Indeed, computing the dimensionless quantities used in \cite{Schmidt97}, we get
$h\simeq 0.56$ and $\rho_H \simeq 0.64$, which corresponds to the onset 
of crystallisation, arriving from the fluid sector.
Furthermore, for hard core radius $\dhc\sqrt{\sigma} \lessapprox 0.4$ (or equivalently, $r\lessapprox 0.5$), only minor differences are seen compared to the $\dhc=0$ situation in this low-coupling regime, where electrostatics can be described in a mean-field manner. We will see below when discussing the no-hindrance regime that this
insensitivity is even more pronounced in the strong-coupling regime since it holds strictly for $r<1$, and also in a sense to be specified for $r>1$.

At $\Xi=17.5$, and at short separations, one observes a shallow attraction between the two surfaces if the counterions radius is not too large, $\widetilde{d}_{\rm hc}<5.25$ (or $\dhc\sqrt{\sigma} <0.5$). Pressure curves start to be influenced by hard core radius as soon as $\widetilde{d}_{\rm hc} \gtrapprox 4$. Increasing counterions size makes pressure curves repulsive at all separations for $\widetilde{d}_{\rm hc} \approx 5.25$, but with local minima for $r<1$ ($\widetilde{d}_{\rm hc}<7.98$). The peak at $\eta\simeq0.5$ seen previously for low couplings and $\dhc\sqrt{\sigma}=0.7$ appears also for the lower $\dhc\sqrt{\sigma}$ values (0.5 and 0.6) at $\Xi=158$ to gradually disappear again at even higher coupling parameters. $\Xi=17.5$ is special in the sense that pressures are close to zero around $\eta=0.5$ for the point charge case and hence is sensitive for perturbations (\emph{e.g.} introducing excluded volume) around this state. 
The local minimum seen for $\dhc\sqrt{\sigma}=1$ for the low coupling limit persists up to $\Xi=17.5$, but vanishes somewhere in the range $\Xi=[17.5,158]$. Even though the pressures are in practice zero for the high coupling cases at $\eta=1.5$ we still do not see any effect of the hard core radius (in contradiction to the $\Xi=17.5$ case and states around $\eta=0.5$).
The relative influence of hard core vs electrostatic interactions is also decreasing with increasing coupling parameter. By increasing $\Xi$, one turns the $r<1$ cases from repulsive at all separations to attractive, except in a narrow range close to zero separation, of extension given by the Gouy-Chapman length. Somewhere around $\Xi=1750$, one also turns cases with $r>1$ from purely repulsive to having an attractive pressure minimum. For our highest coupling parameter ($\Xi=175000$), we see that the $r\geq 1$ follows the $\dhc=0$ curve up to the closest separation for the corresponding $r$. This can be viewed as an extension to the sector $r>1$ of the no-hindrance effect
alluded to in section \ref{Sec3}: it indicates that under such strong couplings, the dominant effect is electrostatics, equivalent to that of point-like ions, while steric effects only matter through the forbidden overlaps. When no overlaps are involved, the Coulombic interactions are largely dominant.
The case $r=1$ does, however, have a smaller minimum in absolute value compared to the $\dhc=0$ case even though the closest approaches are the same. This can theoretical be understood as some of the preferred bilayer structures are forbidden due to hard core overlaps (see Fig. \ref{fig:etamin_Delta_comment}), leading to a slightly altered and weaker (in terms of attraction) pressure curve.
We come back to this in the next subsection.

\begin{figure*}[htbp]
\includegraphics[width=0.325\textwidth,clip]{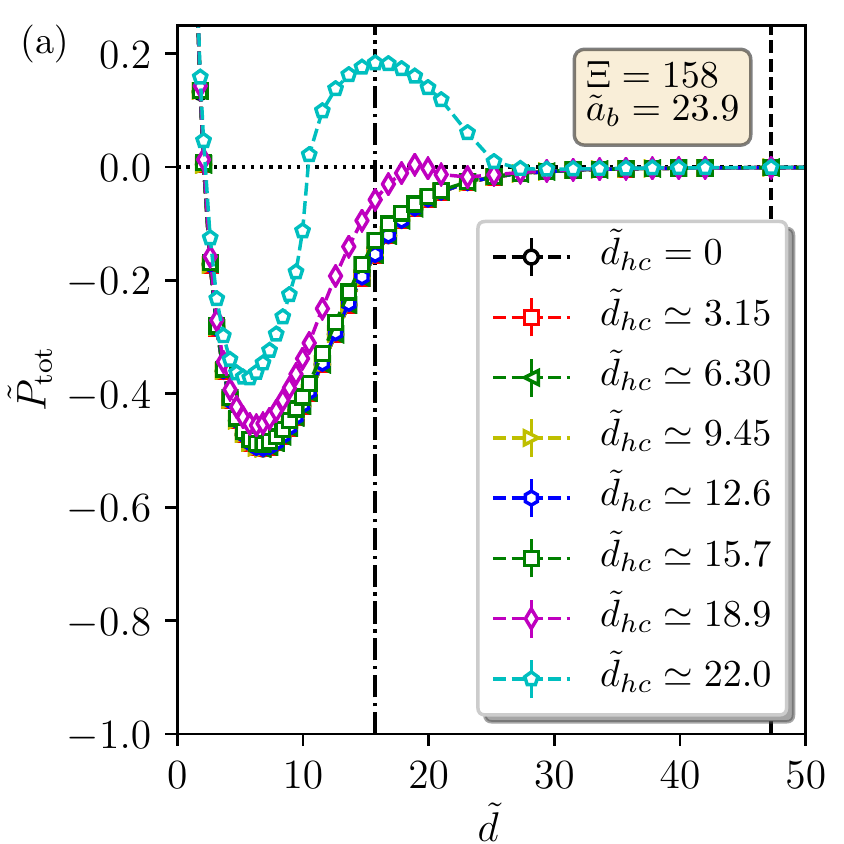}
\includegraphics[width=0.325\textwidth,clip]{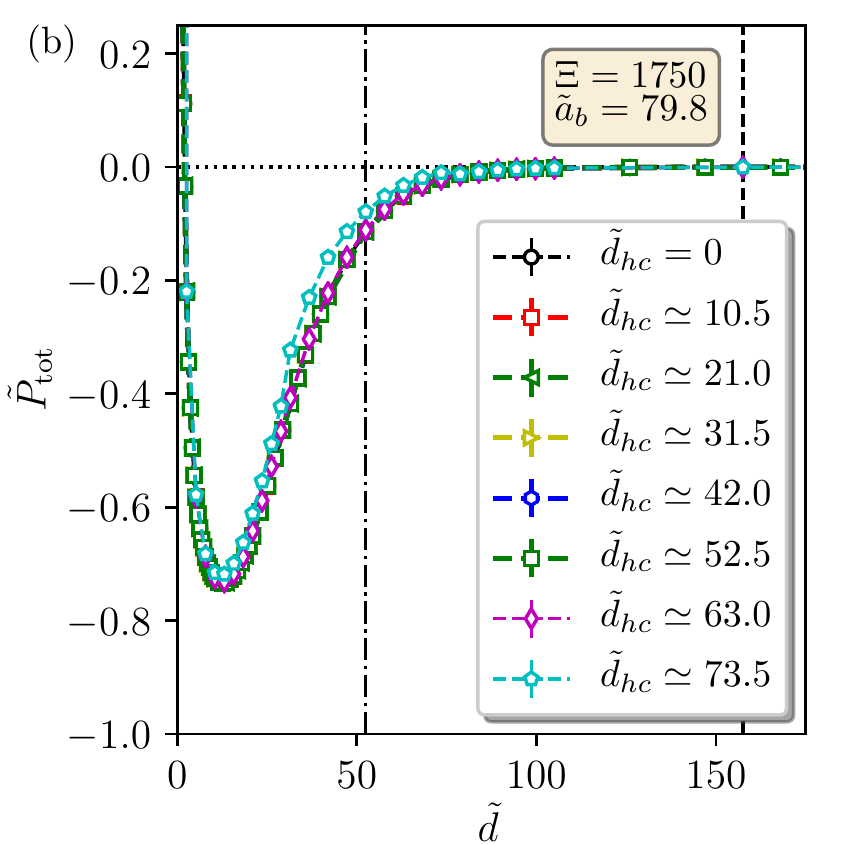}
\includegraphics[width=0.325\textwidth,clip]{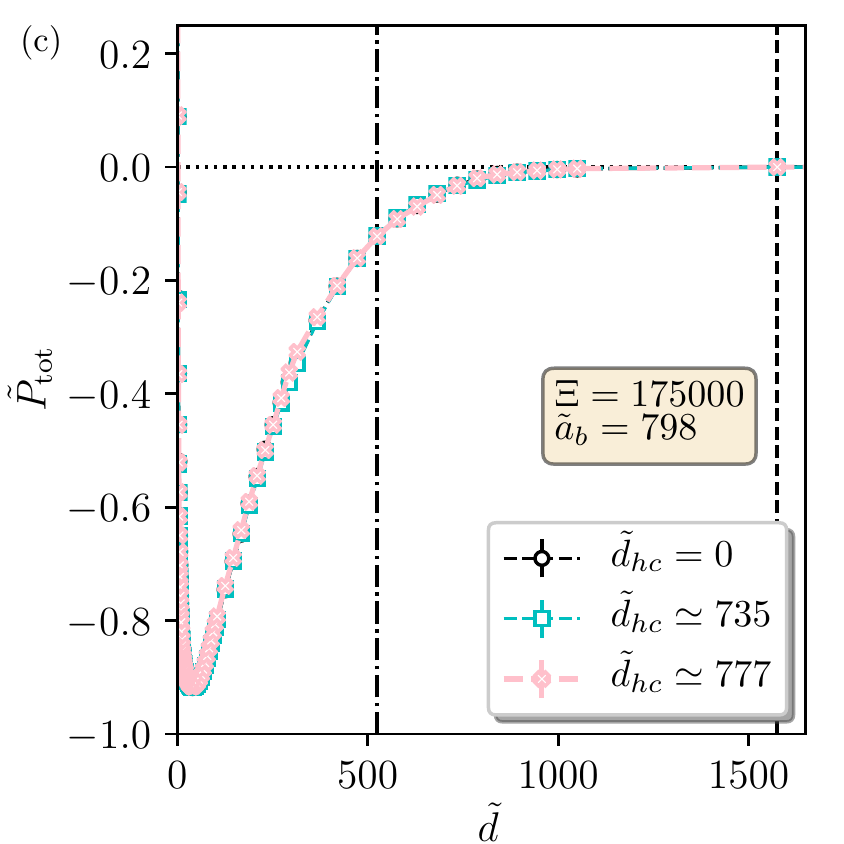}
\caption{Monte Carlo equation of state in the $r<1$ regime, i.e. when a monolayer fits in between the two plates at closest separation ($d=0$). 
(a) $\Xi=158$, (b) $\Xi=1750$, and (c) $\Xi=175000$.
Due to the strong coulombic in-plane repulsion in panels (b) and (c), 
steric effects hardly affect the pressure, nor the density profiles, beyond the trivial shift of closest distance ($D \to D-\dhc = d$).
  In all figures symbols and colors correspond to $\dhc \sqrt{\sigma}$= 0 (black circles), 0.1 (red squares), 0.2 (green triangles pointing left), 0.3 (yellow triangles pointing right), 0.4 (blue hexagons), 0.5 (green squares), 0.6 (maroon diamonds), 0.7 (cyan pentagons), and 0.74 (pink crosses). In panel c), all values of $\widetilde d_{\rm hc}$ below 735 also lead to excellent collapse onto
the  $\widetilde d_{\rm hc}=0$ curve.}
\label{fig:Pressure_noHindrance}
\end{figure*}

\subsection{Comparison with analytic results}
\label{ssec:comparison}

\begin{figure}[htb]
\includegraphics[width=0.325\textwidth,clip]{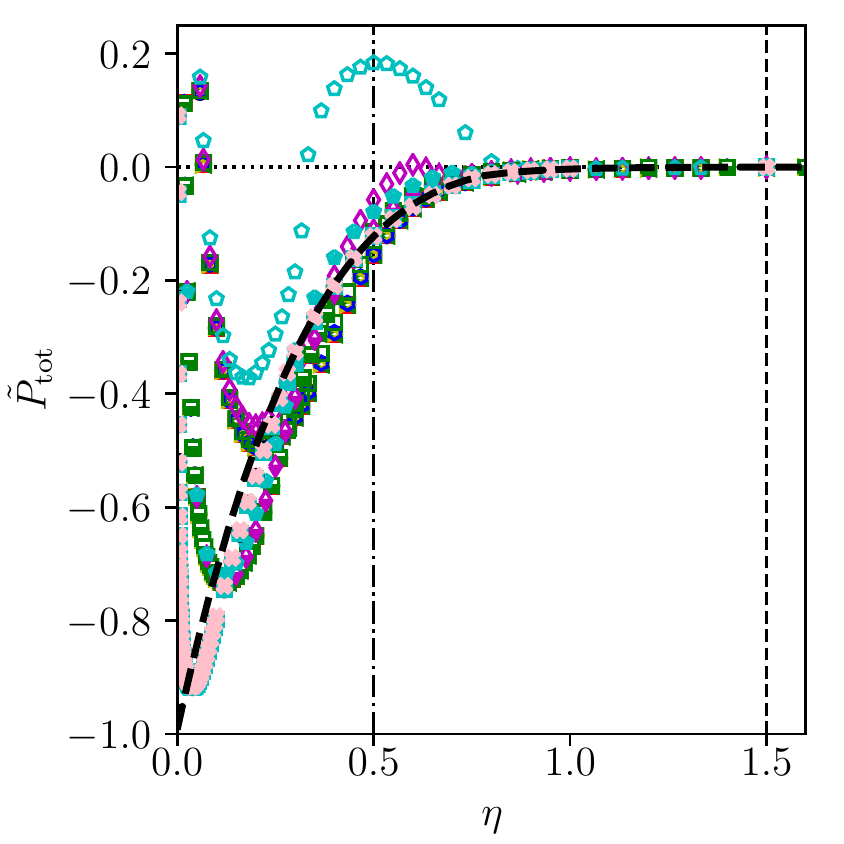}
\caption{Same as Fig \ref{fig:Pressure_noHindrance}, in a difference scale for interplate distance, with $\eta$ rather than $\widetilde d$. The increasing branches of the pressure do collapse on a master curve (dashed line), that corresponds to the ground-state electrostatic pressure for point ions, realised when $\Xi\to \infty$ (formally speaking, at vanishing temperature \cite{Samaj12a,Ivan,Palaia20}). Indeed, the dashed line displays the corresponding force per unit surface.
Symbols and colors as in Fig. \ref{fig:Pressure_noHindrance} with open markers for $\Xi=158$, half-filled for $\Xi=1750$, and full for $\Xi=175000$.
}
\label{fig:Pincrease_universal}
\end{figure}

\begin{figure}[htb]
\includegraphics[width=0.325\textwidth,clip]{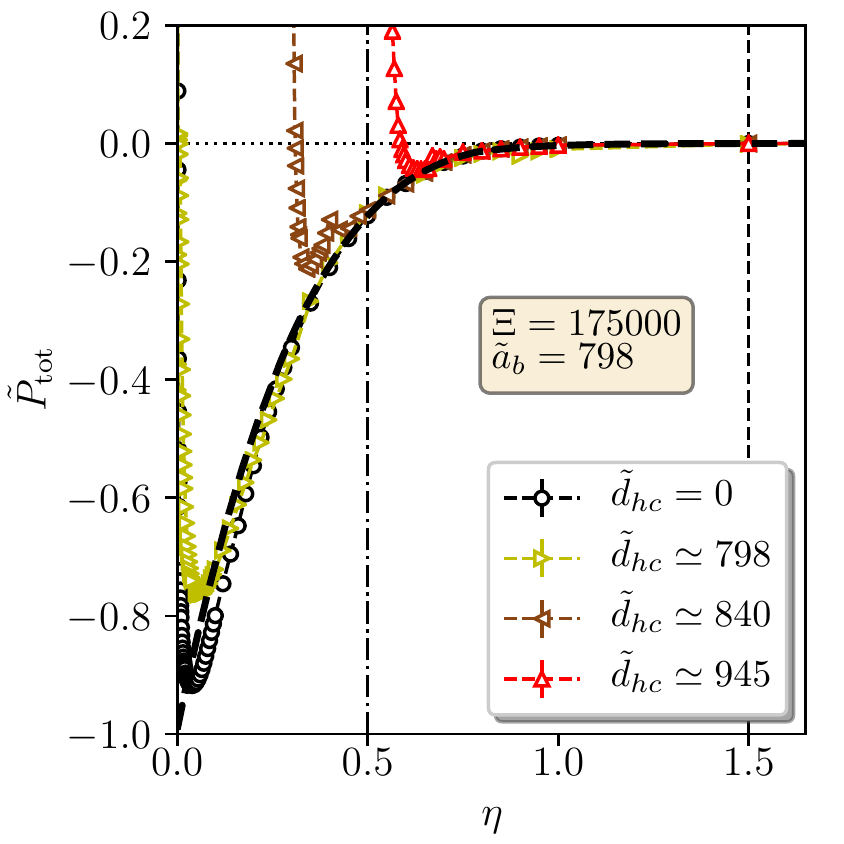}
\caption{Same as Fig \ref{fig:Pincrease_universal}, for $\Xi=175000$, with $r=1$ ($\dhc\sqrt\sigma=0.76$ and $\widetilde{d}_{\rm hc}=798$),
$r=1.05$ ($\dhc\sqrt\sigma=0.8$ and $\widetilde{d}_{\rm hc}=840$), and
$r=1.18$ ($\dhc\sqrt\sigma=0.9$ and $\widetilde{d}_{\rm hc}=945$).
}
\label{fig:Pincrease_universal_rbig}
\end{figure}

We have argued in section \ref{Sec3} that for $\dhc < a_b$, all pressure curves
$P(d)$ should collapse onto their point-ion limit, provided the coupling parameter $\Xi$ be large enough. This no-hindrance regime is illustrated in Fig. \ref{fig:Pressure_noHindrance}. In this figure, the largest value 
of $r$ reported (respectively 0.92, 0.92 and 0.97 for panels a), b) and c))
is fairly close to 1. Yet, at the largest $\Xi$, this has no visible effect 
on the pressure curve, while the quality of the data collapse is altered 
when decreasing coupling $\Xi$.
In the corresponding equation of state
the increasing branch on the right hand side is actually universal, independent
of coupling $\Xi$, when expressed in the proper variable, here $\eta$ \cite{Ivan,Palaia20},
as revealed in Fig. \ref{fig:Pincrease_universal}. The reason behind this universality is that the behaviour is ruled by the infinite coupling attractor of point ions
(with a divergent $\Xi$). The point-ion ground state pressure is indeed shown by the dashed line in Fig. \ref{fig:Pincrease_universal}.
Besides, as hinted at in subsection \ref{ssec:MC}, the no-hindrance effect extends to
case with $r>1$, see Fig. \ref{fig:Pincrease_universal_rbig}. For 
$\dhc > a_b$, i.e. $r>1$, it is seen that starting from large distances, 
the pressure curve follows the point counterion limiting curve, down to the smallest distance allowed by non overlap of hard cores. The minimal distance
given by Eq. \eqref{etamin}, matches very well that where $\widetilde P_{\rm tot}$ diverges in 
Fig. \ref{fig:Pincrease_universal_rbig}: for the data with $r\simeq 1.05$,
we have $\eta_{\min}(\sqrt{3}/r^2,r) \simeq 0.29$ (see also Fig. \ref{fig:etamin_Delta}), while for $r\simeq 1.18$, we get $\eta_{\min}(\sqrt{3}/r^2,r) \simeq 0.54$. Steric effects are here dichotomic: there are essentially irrelevant due to the strong Coulombic repulsion,
or divergent at small $d$, for those configurations which are not allowed.
Quite remarkably, the marginal situation with $r=1$ remains close to 
the point-ion attractor, down to vanishing distances where the two double layers on the opposite walls exactly register.

\begin{figure}[htb]
\includegraphics[width=0.325\textwidth,clip]{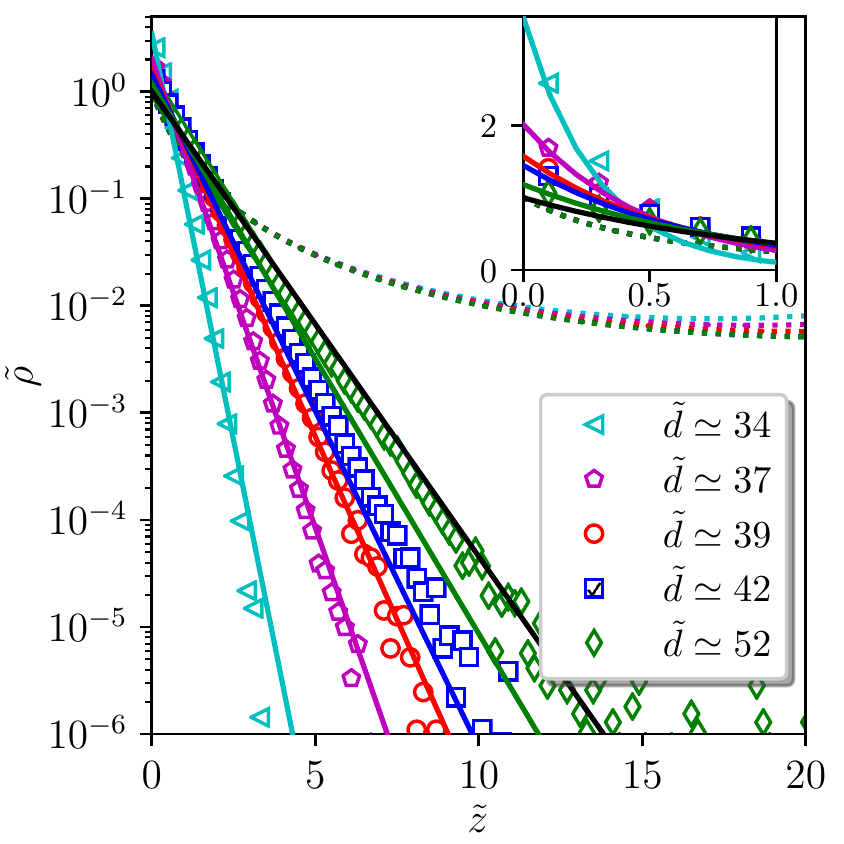}
\caption{Ionic density profile, in linear-log scale (linear-linear in inset), for $\Xi=1750$ and $\widetilde{d}_{\rm hc}=84.0$ at various separations $\widetilde{d}$.
The black line has slope 1 which corresponds to the strong-coupling
one-wall case. 
The other lines are guides to the eye,
from which we extract the slope reported in the inset of Fig. \ref{fig:alpha}. The dashed lines, which are almost superimposed, 
are density profiles obtained from the Poisson-Boltzmann equation for the same separations (colors according to separation).
}
\label{fig:profiles}
\end{figure}

\begin{figure*}[htb]
\includegraphics[width=0.325\textwidth,clip]{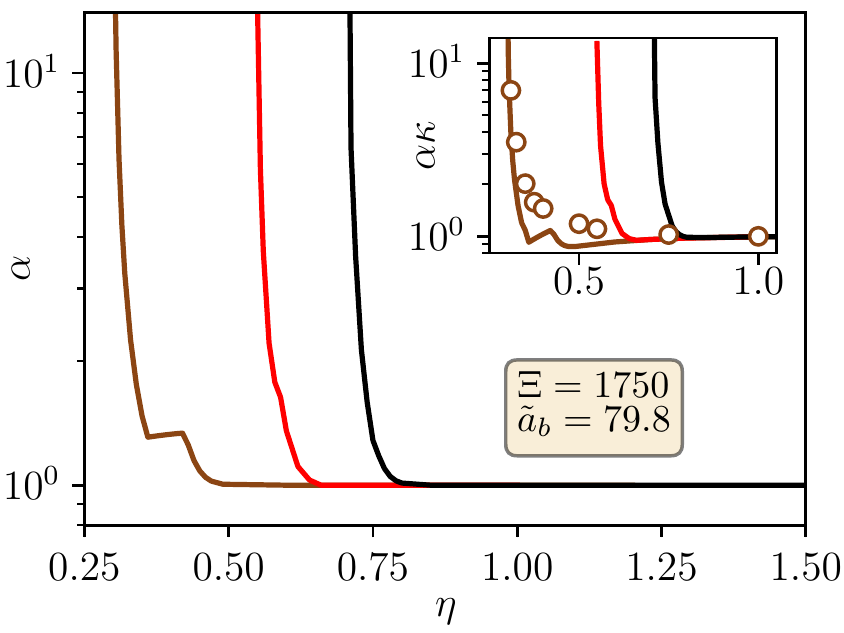}
\includegraphics[width=0.325\textwidth,clip]{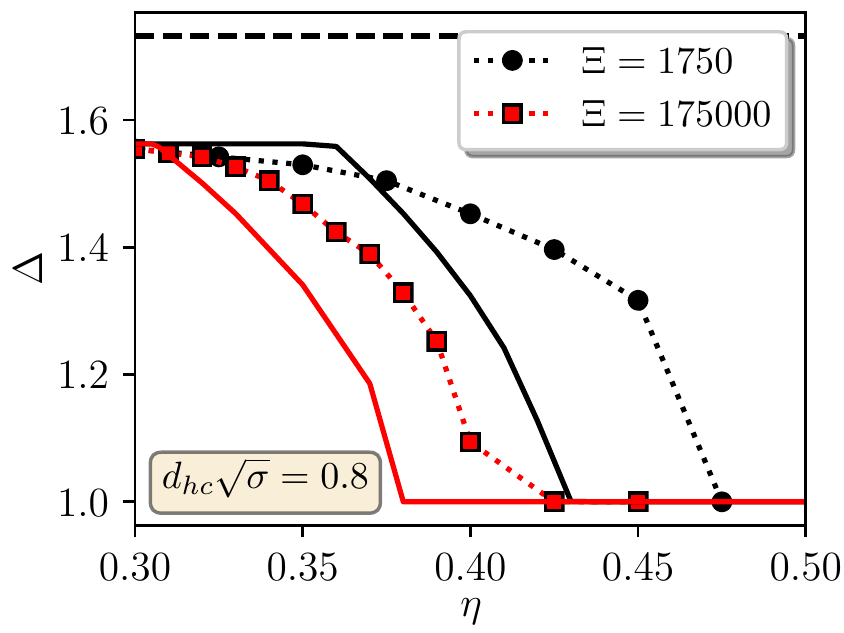}
\caption{Left: Evolution of the localization parameter $\alpha$, as entering \eqref{eq:p0ansatz} and \eqref{eq:alpha_rho}, with distance $\eta$ for $\Xi=1750$. From left to right, the curves correspond to 
$\widetilde{d}_{\rm hc}=84$, $\widetilde{d}_{\rm hc}=94.5$ and $\widetilde{d}_{\rm hc}=105$.
The product $\alpha \kappa$, shown in the inset, is the local electric field felt by an ion in the vicinity of a given plate. 
This quantity embodies the influence of steric effects on this field, which does rule the ionic profiles, see Eqs. \eqref{eq:alpha_rho1} and \eqref{eq:alpha_rho}.
In the inset, the symbols are for the MC measures, as extracted from data like those
presented in Fig. \ref{fig:profiles}, for $\widetilde{d}_{\rm hc}=84$.
Right: Corresponding dependence of the aspect ration $\Delta$ with distance, for $ d_{\rm hc}\sqrt{\sigma}=0.8$, meaning  $r\simeq 1.05 $, 
so that $\widetilde{d}_{\rm hc} = 84$ at $\Xi=1750$, and $\widetilde{d}_{\rm hc} = 840$ at $\Xi=175000$. The full line is for the theoretical prediction and the 
symbols correspond to the MC measures (dotted lines are guides to the eye). The dashed horizontal line is at $\sqrt{3}$.
}
\label{fig:alpha}
\end{figure*}

\begin{figure*}[htb]
\includegraphics[width=0.325\textwidth,clip]{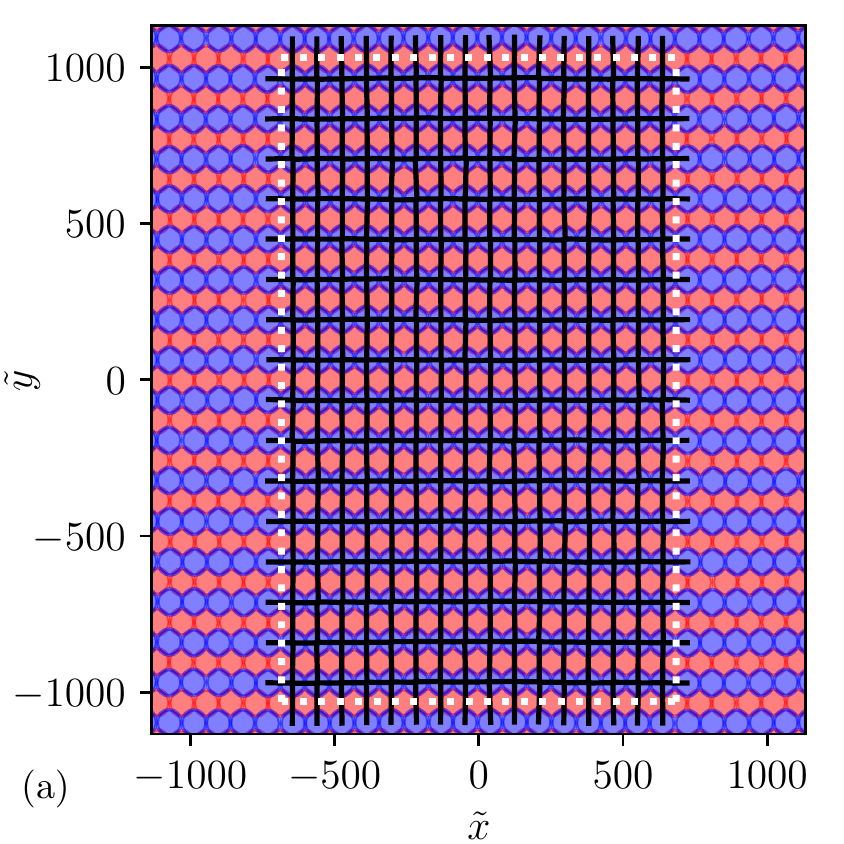}
\includegraphics[width=0.325\textwidth,clip]{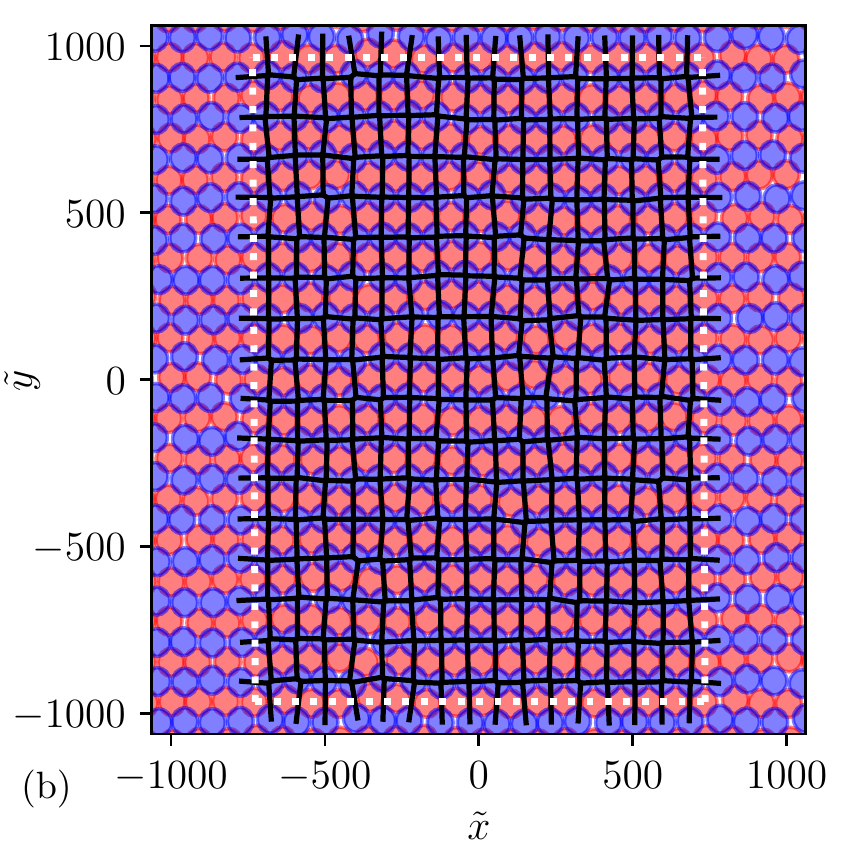}
\includegraphics[width=0.325\textwidth,clip]{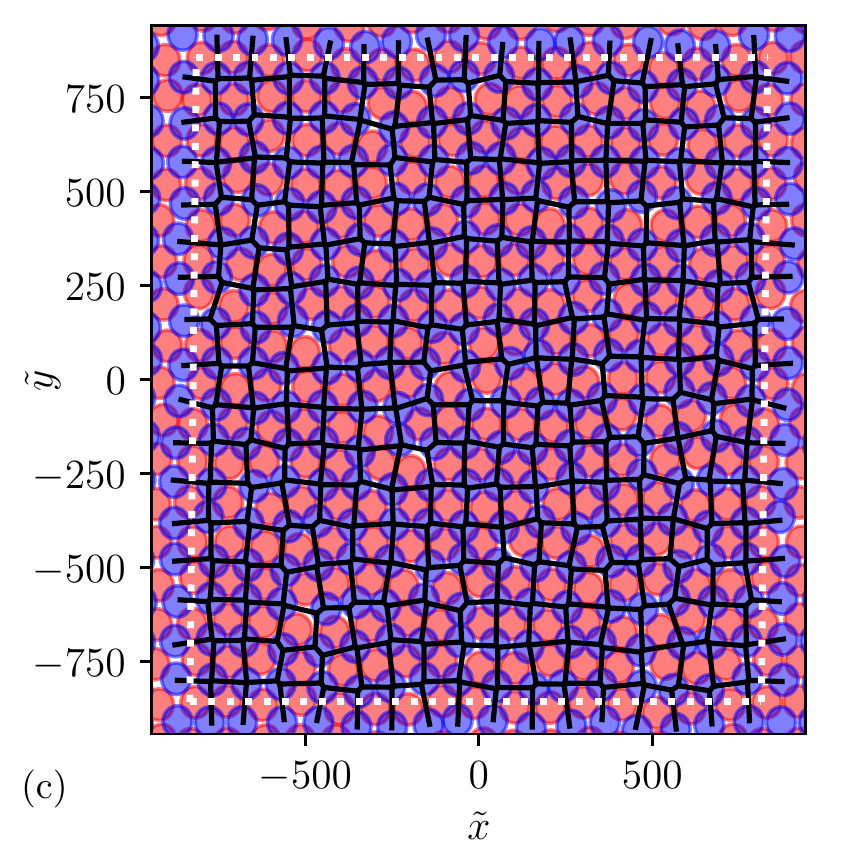}
\caption{Evolution of the structures (2D projection of instantanous configurations onto the $x$-$y$ plane) at $\Xi=1750$ and $\widetilde{d}_{hc}=84$ when increasing interplate separations. (a) $\eta=0.35$, (b) $\eta=0.425$, and (c) $\eta=0.5$. 
Colors are assigned depending on the closest wall for each particle. White dashed lines indicate the main simulation box and black lines correspond to a Voronoi construction for the red particles (say those on plate 1, with neighbors on plate 1). Size of the particles corresponds to the hard-core. Notice that structures become liquid above $\eta\simeq 0.6$.
}
\label{fig:snapshots}
\end{figure*}

\begin{figure*}[htb]
\includegraphics[width=0.325\textwidth,clip]{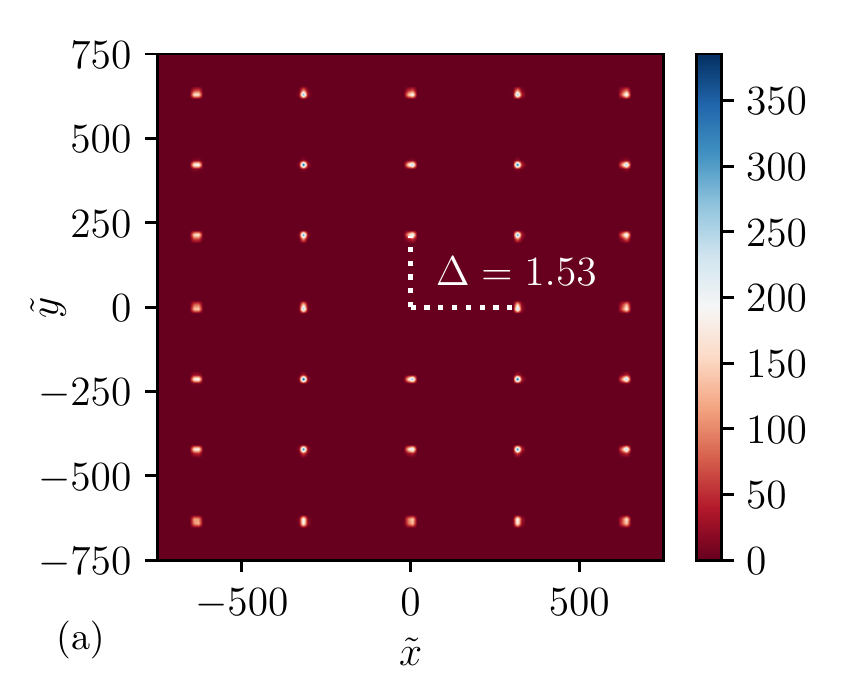}
\includegraphics[width=0.325\textwidth,clip]{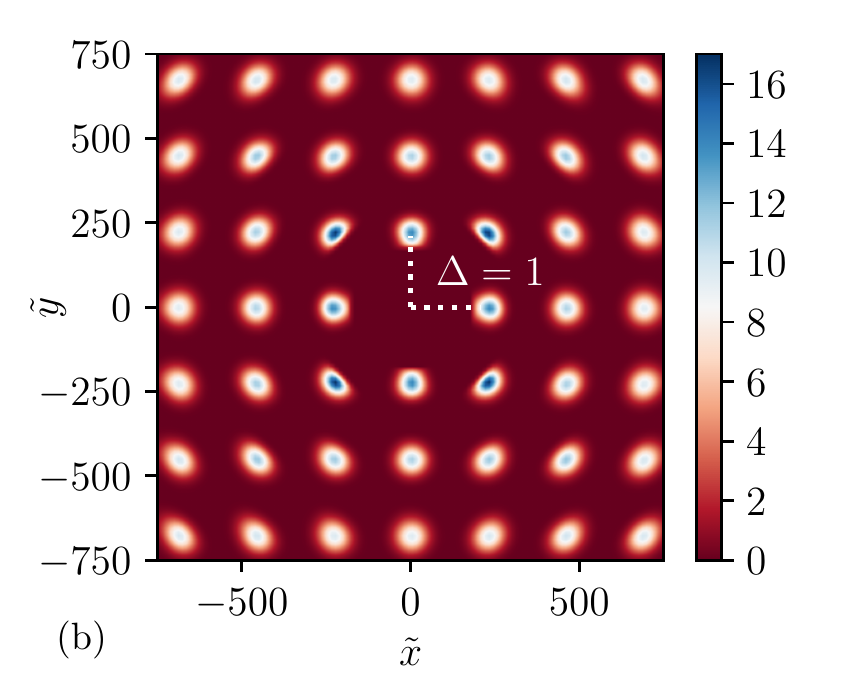}
\includegraphics[width=0.325\textwidth,clip]{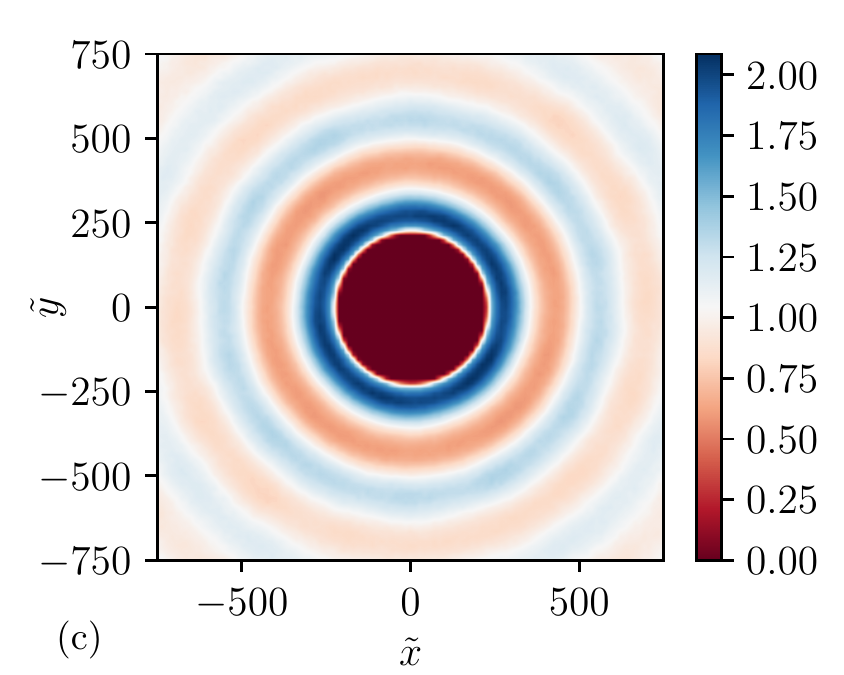}
\includegraphics[width=0.325\textwidth,clip]{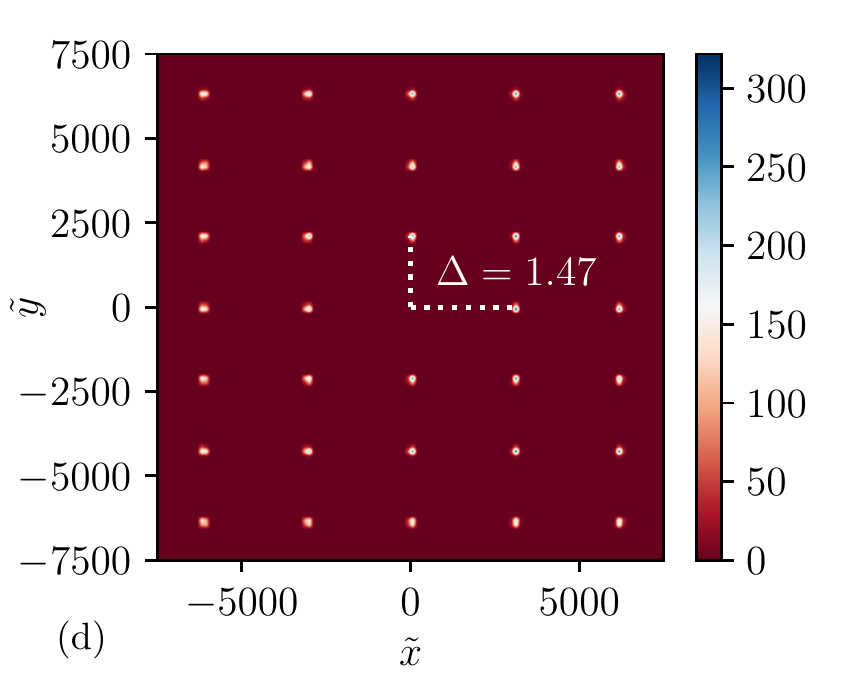}
\includegraphics[width=0.325\textwidth,clip]{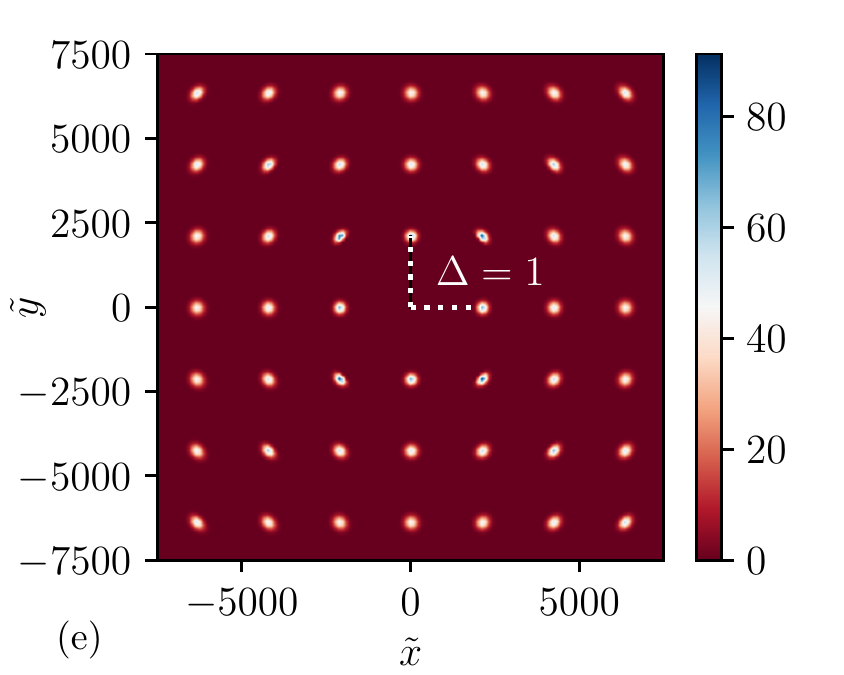}
\includegraphics[width=0.325\textwidth,clip]{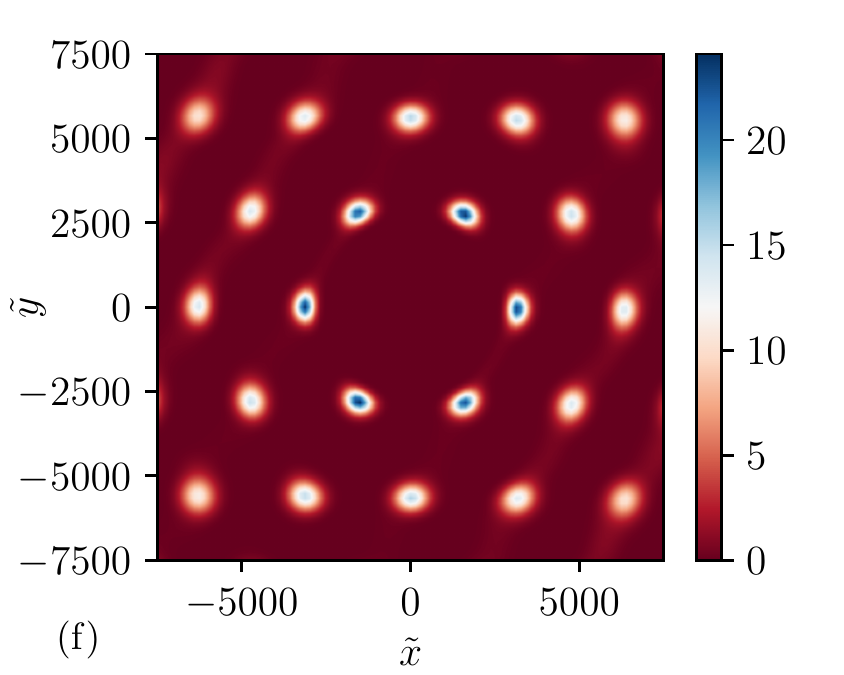}
\caption{Evolution of the 2D pair-correlation functions $g_{2d}(x,y)$ (2D projections of onto the $x$-$y$ plane) of particles on the same side of the midplane with $\Xi=1750$ and $\widetilde{d}_{hc}=84$ for separations (a) $\eta=0.35$, (b) $\eta=0.5$, and (c) $\eta=0.7$. For $\Xi=175000$ at $\widetilde{d}_{hc}=840$ for separations (d) $\eta=0.35$, (e) $\eta=0.425$, and (f) $\eta=0.8$. White dotted lines illustrates the Wigner cell dimensions and the $\Delta$-value. Note that one obtains a liquid structure above $\eta\simeq 0.6$ for $\Xi=1750$ while for $\Xi=175000$ one obtains a hexagonal monolayer (so-called structure V \cite{Samaj12a}). A transition between structure III and V occurs between $\eta=0.6$ and $\eta=0.75$ (data not shown); under infinite coupling, structure V becomes stable for $\eta>0.73$ \cite{Samaj12a}.
}
\label{fig:2D_gr}
\end{figure*}

Before entering into a more precise comparison between analytical and numerical pressure profiles, we test the ansatz underlying our choice of trial exponential-type density $p_0$
in Eq. \eqref{eq:p0ansatz}, by showing the ionic profiles in Fig. 
\ref{fig:profiles}. The first observation is that they are neatly exponential
in the vicinity of the plates, in line with our premises for the trial 
variational form for $p_0$.
The profiles significantly deviate from their mean-field, Poisson-Boltzmann counterpart, shown with the dashed lines;
on the scale of the figure, the different dashed lines corresponding
to different separations are quasi-superimposed, but would depart at larger distances.
We have predicted in section \ref{Sec3}
that steric effects make the ionic profiles more peaked at the plates than the equivalent point-ion system, having the same 
available free space for center-of-mass displacement. 
This is corroborated by the MC data.
This steric-driven enhanced localization is quantified, in the theory,
by the parameter $\alpha$ appearing in \eqref{eq:p0ansatz} and \eqref{eq:alpha_rho}. The predicted behaviour of $\alpha$ is shown in Fig. 
\ref{fig:alpha}-left. When the plates are far away, $\alpha\to 1$, signalling 
that steric effect do not affect the local field $\kappa$, itself distance dependent, that maintains ions in the vicinity of a given plate. On the other hand, decreasing $\eta$, it is seen that
$\alpha$ increases quite significantly. 
Besides, it is the product $\alpha \kappa$ that defines the local field; $\kappa$ decreases for decreasing $\eta$ while $\alpha$ shows the opposite trend, and we show in the inset of Fig. \ref{fig:alpha}-left the resulting 
effects for the product, compared to the MC measures. 
These MC results are extracted from the slopes evidenced in Fig. 
\ref{fig:profiles}.
While the predicted trend seems correct, it is seen that the theory leads to too sharp of a dependence
on the distance, while MC yields smoother curves. 
A similar comment applies to the aspect ratio parameter, displayed in 
Fig. \ref{fig:alpha}-right. Our variational treatment captures the correct trend, but exaggerates the sharpness of the crossover. 
We note nevertheless that the agreement between MC and the prediction improves,
expectedly,
when increasing $\Xi$.
    It can be noted that the distance range where 
the predicted $\Delta$ underestimates the measured one is precisely 
where the localisation parameter $\alpha$ in Fig. \ref{fig:alpha}-left
displays a non-monotonous local bump. 
It is also worthwhile here to inspect directly structural features.
Fig. \ref{fig:snapshots} shows the projected instantaneous  
position of ions, which reveals that the arrangement is of type 
considered in the theoretical analysis, with rectangular unit cells. 
From left to right, the aspect ratio $\Delta$ decreases from 1.53
 to a value close to 1, as also shown in Fig. 
 \ref{fig:alpha}-right. This is confirmed by the computation 
 of the in-plane pair correlation function, as displayed in Fig. 
 \ref{fig:2D_gr}, which also illustrates the relevance of $\Xi$ to 
 maintain in-plane order at large separation.

\begin{figure*}[t!]
\includegraphics[width=0.325\textwidth,clip]{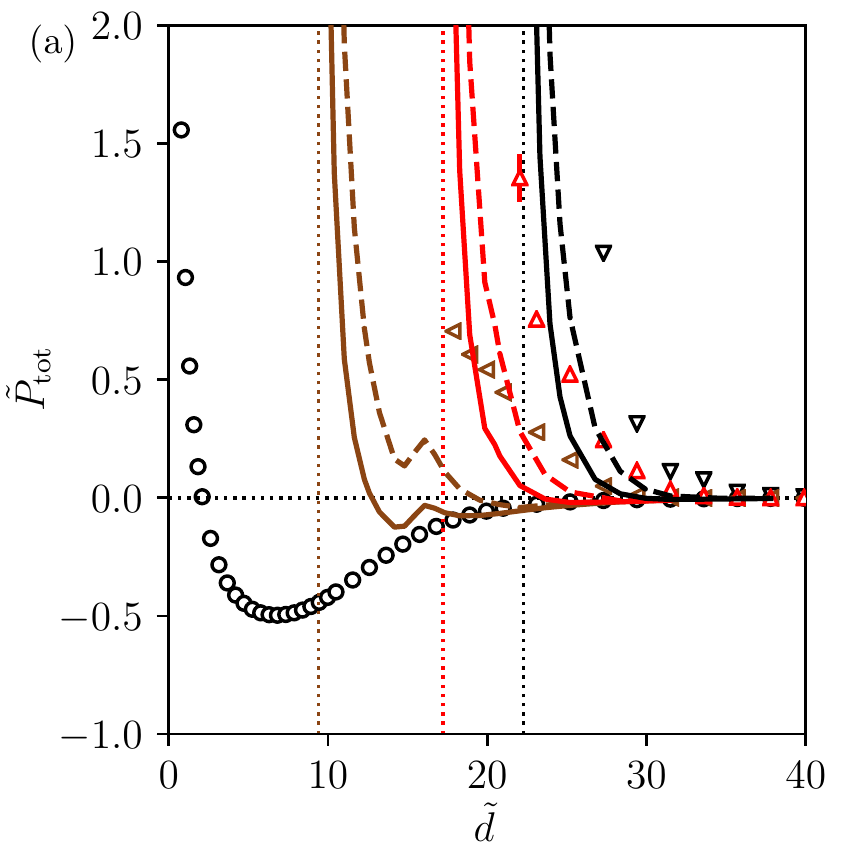}
\includegraphics[width=0.325\textwidth,clip]{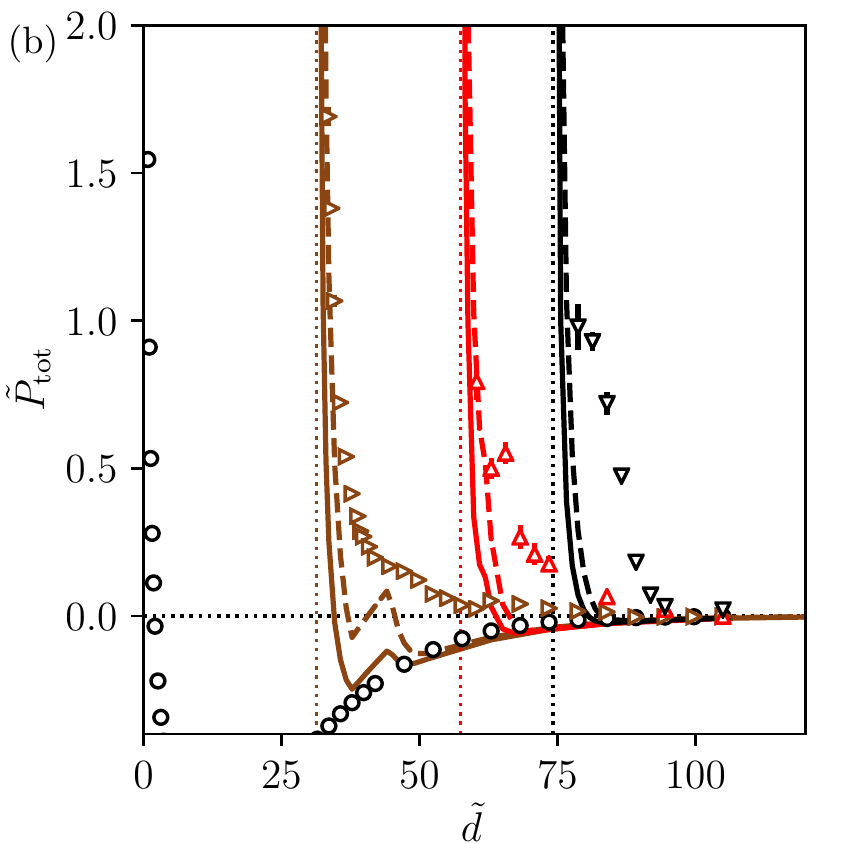}
\includegraphics[width=0.325\textwidth,clip]{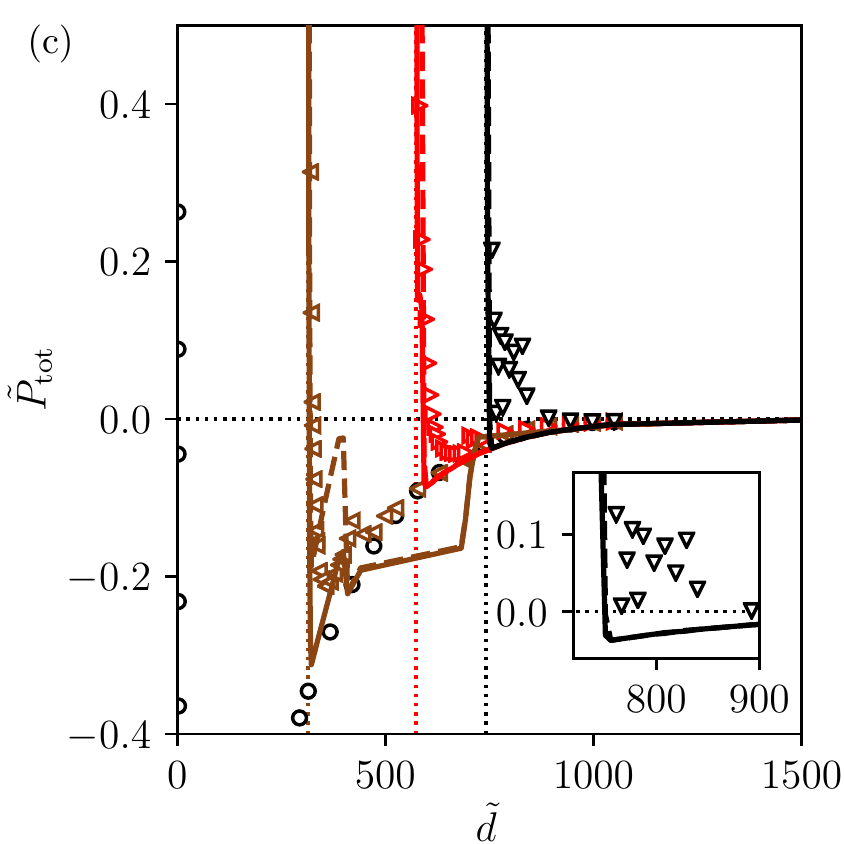}
\caption{Analytical equation of states compared to Monte Carlo results for: (a) $\Xi=158$, (b) $\Xi=1750$, and (c) $\Xi=175000$. Symbols correspond to Monte Carlo results with the same labeling as in Fig. \ref{fig:Pressure_MC}. The lines show the analytic predictions using the thermodynamic (solid) and the contact route (dashed).
Vertical dotted lines show the respective closest approach distances,
as given by Eq. \eqref{etamin}. The inset in (c) shows the numerical noise of the Monte Carlo data together with the analytical prediction.}
\label{fig:Pressure_comp}
\end{figure*}

Figure \ref{fig:Pressure_comp} shows a comparison of the pressures obtained from the analytical theory and numerical Monte Carlo simulations at coupling parameters $\Xi=158$, 1750, 175 000 and for various hard core radii of the counterions. We focus here on the $r>1$ cases. For $\Xi=158$, the analytic theory seems to underestimate the repulsive pressure due to the hard core (or equivalent overestimate the electrostatic attraction). Similar trend is seen at $\Xi=1750$ even though the theory yields pressures closer to the numerical results. For $\Xi=175000$, we find a good agreement between numerical results and theory. At $\dhc\sqrt{\sigma}=1$ ($\widetilde{d}_{hc}=1050$, i.e. $r=3^{1/4}$), the numerical Monte Carlo data does however exhibit a significant level of noise. Besides, we have proposed two routes to compute the pressures,
a mechanical and a thermodynamical one. Within an exact treatment, both results
should coincide. The fact that they yield relatively close results in Fig. \ref{fig:Pressure_comp} assesses 
the self-consistency of the approach proposed.


To summarize, the agreement between our analytical calculations and the Monte Carlo results is good at the highest coupling studied, where steric effects
either do not alter the point-ion pressure, or forbid too close interplate distances, for which the pressure is infinite. Steric effects are thus here dichotomic, all or nothing. 
At smaller couplings, a crossover sets in, where hard core have a finite 
and non negligible contribution to the pressure, that we capture semi-quantitatively,
see Fig. \ref{fig:Pressure_comp}. Difference between the theory and simulation data 
are due to the mean-field nature of the variational prediction performed.

\section{Conclusion} \label{Sec5}
We have derived an analytic strong-coupling theory for two like-charged plates, treating counterions as charged hard spheres of diameter $\dhc$. Coulombic coupling is measured by a parameter $\Xi$ that weights electrostatic effects against thermal energy. Starting from point charges and increasing $\dhc$, a regime appears where it is no longer  possible to accommodate a layer of counterions between the plates, but a bilayer forms. At this point (corresponding to a ratio of $\dhc$ over lattice spacing $r=1$), the distant plates nevertheless can accommodate a monolayer of counterions, under larger coulombic couplings. We did not treat the cases of still larger values of $\dhc$, where steric repulsion would lead to
more complex arrangements (multilayers), in particular for $r>3^{1/4}$. Our approach starts from crystalline configuration 
of counterions that form at large $\Xi$. These crystals, which are staggered
between the two plates, have been assumed to 
have a rectangular unit cell, which, given the staggering, includes the triangular lattice (often referred to as ``hexagonal'') that forms at close contact when a monolayer is admissible \cite{rque100}. By integrating out all ions in the vicinity of plate 2, we obtain a non-trivial effective Hamiltonian ruling 
the behaviour of ions in the vicinity of plate 1. For the sake of tractability,
an upper bound to the corresponding free energy is computed in the Gibbs-Bogoliubov spirit, considering a family of factorized probability distribution for the ions positions that involve a localisation parameter $\alpha$. By minimizing this bound with respect to $\alpha$ and the lattice aspect ratio, we obtain explicit density profiles and pressures.

Our predictions have been compared to Monte Carlo simulations.
At the largest $\Xi$, the results show a remarkable insensitivity to 
hard-core diameter, not only for $r<1$ but also above 1, provided one works with the shifted distance between the plates ($D \to D-\dhc$). 
Only in the small distance range that is ruled out due to unavoidable ionic overlaps is the point-like pressure
inapplicable. Decreasing $\Xi$, packing effects prove more relevant and have a non trivial signature on the equation of state, that our theory captures
in a semi-quantitative way (see Fig. \ref{fig:Pressure_comp}). We also found numerically that steric effects can completely suppress like-charge attraction \remma{within the primitive model},
although less efficiently when $\Xi$ increases. Indeed, the larger the coupling,
the more ions do repel, and the less relevant their hard core becomes. \remma{We believe that this effect (\emph{e.g.} suppression of like-like attraction) still persists in the low salt concentration cases and that steric effects in general will become more important as the salt concentration is further increased (\emph{i.e.} increased repulsion in the pressure curves), as has been seen for the Debye screening length \cite{Smith16}. } 

This work paves the way towards a more satisfactory and realistic description of strong-coupling theory, beyond the point ion limit for which it was initially devised.
Among interesting perspectives, we mention the study of larger $r$ values,
when the ionic diameter would lead to multi-layers at close packing,
when the two plates are at closest separation \remma{or the effect of the size and structure of the solvent itself}. It would also be relevant to address asymmetrically charged walls, together with systems with salt,
when micro-ions of both signs are present, not only counterions. \remma{In this respect, a promising
approach is to extend the analysis of \cite{Paillusson11}, where ions of opposite charges 
form Bjerrum pairs, {\em i.e}. neutral entities
that may be in a first approximation discarded from the analysis. This yields an effective salt-free system,
as addressed in the present work.}
Besides, in severely confined configurations, the modification of the solvent (say water) dielectric constant should also be included in the description
\cite{Mukina19,Schlaich19}.

We would like to thank I. Palaia and J. Zelko for useful discussions. The support of L. \v{S} received from the project EXSES APVV-16-0186 and VEGA Grant 2/0003/18 is aknowledged.

\renewcommand{\theequation}{A\arabic{equation}}
\setcounter{equation}{0}

\begin{appendix}
\section{Generalized Misra functions} 
\label{app:A}
The first few generalized Misra functions $z_{\nu}(x,y)$ (\ref{znu}) with 
half-integer arguments are expressible in terms of the complementary error 
function \cite{Gradshteyn}
\begin{equation} \label{erfc}
{\rm erfc}(u)=\frac{2}{\sqrt{\pi}}\int_u^\infty\exp{(-t^2)}\ {\rm d}t ,
\end{equation}
as follows \cite{Travenec15}:
\begin{widetext}
\begin{eqnarray} 
z_{1/2}(x,y) & = & \sqrt{\frac{\pi}{x}}{\rm e}^{-2\sqrt{xy}}
\left[ 1-\frac{1}{2}\ {\rm erfc}{\left(\sqrt{\frac{x}{\pi}}-\sqrt{\pi y}\right)}
-\frac{1}{2}{\rm e}^{4\sqrt{xy}}\ 
{\rm erfc}{\left(\sqrt{\frac{x}{\pi}}+\sqrt{\pi y}\right)}\right], 
\nonumber \\
z_{3/2}(x,y) & = & \sqrt{\frac{\pi}{y}}{\rm e}^{-2\sqrt{xy}}\left[ 1-\frac{1}{2}\ 
{\rm erfc}{\left(\sqrt{\frac{x}{\pi}}-\sqrt{\pi y}\right)}
+ \frac{1}{2}{\rm e}^{4\sqrt{xy}}\ 
{\rm erfc}{\left(\sqrt{\frac{x}{\pi}}+\sqrt{\pi y}\right)}\right], 
\nonumber \\
z_{5/2}(x,y) & = & \frac{\sqrt{\pi x}}{y}{\rm e}^{-2\sqrt{xy}}
\left(1+\frac{1}{2\sqrt{xy}}\right)
-\frac{\sqrt{\pi}}{4y^{3/2}}\bigg[-4{\rm e}^{-x/\pi-\pi y}\sqrt{y} 
\nonumber \\ & & 
+{\rm e}^{-2\sqrt{xy}}\left(1+2\sqrt{xy}\right)\ 
{\rm erfc}{\left(\sqrt{\frac{x}{\pi}}-\sqrt{\pi y}\right)}
+{\rm e}^{2\sqrt{xy}}\left(-1+2\sqrt{xy}\right)\ 
{\rm erfc}{\left(\sqrt{\frac{x}{\pi}}+\sqrt{\pi y}\right)} \bigg] . 
\label{zerror}
\end{eqnarray}
\end{widetext}
The case of the ordinary Misra functions $z_{\nu}(0,y)$ \cite{Misra} should be 
understood in the sense of the limit $x\to 0$,
\begin{eqnarray}
z_{1/2}(0,y) & = &\frac{2}{\sqrt{\pi}}\left[{\rm e}^{-\pi y} -\pi \sqrt{y} \ 
{\rm erfc}{\left(\sqrt{\pi y}\right)}\right], \nonumber\\
z_{3/2}(0,y) & = &\sqrt{\frac{\pi}{y}}\ {\rm erfc}{\left(\sqrt{\pi y}\right)},
\nonumber\\
z_{5/2}(0,y) & = &\frac{\sqrt{\pi}}{2 y^{3/2}}\left[2 {\rm e}^{-\pi y}\sqrt{y} \ 
+{\rm erfc}{\left(\sqrt{\pi y}\right)}\right] . \label{znu0y}
\end{eqnarray}


The function $\Sigma(\eta,\Delta)$, related to the energy per particle 
for the staggered rectangular bilayers I-III with the aspect ration $\Delta$
via Eq. (\ref{e0}), can be expressed as an infinite series of the generalized
Misra functions (\ref{znu}) as follows
\begin{widetext}
\begin{eqnarray}
\Sigma(\eta,\Delta) & = & 4 \sum_{j=1}^{\infty} \left[
z_{3/2}\left(0,j^2/\Delta\right) + z_{3/2}\left(0,j^2\Delta\right) \right]
+ 8 \sum_{j,k=1}^{\infty} z_{3/2}\left(0,j^2/\Delta+k^2\Delta\right) \nonumber \\
& & + 2 \sum_{j=1}^{\infty} (-1)^j \left[
z_{3/2}\left((\pi\eta)^2,j^2/\Delta\right) + 
z_{3/2}\left((\pi\eta)^2,j^2\Delta\right) \right] 
+ 4 \sum_{j,k=1}^{\infty} (-1)^j (-1)^k 
z_{3/2}\left((\pi\eta)^2,j^2/\Delta+k^2\Delta\right) \nonumber \\
& & + 4 \sum_{j,k=1}^{\infty} 
z_{3/2}\left(0,\eta^2+(j-1/2)^2/\Delta+(k-1/2)^2\Delta\right) - 4\sqrt{\pi}
-\pi z_{1/2}(0,\eta^2) . \label{sigmaMisra} 
\end{eqnarray}
\end{widetext}

\renewcommand{\theequation}{B\arabic{equation}}
\setcounter{equation}{0}
\section{Calculation of the density profile} 
\label{app:B}
To derive the particle number density, we add to the Hamiltonian for each
particle the generating one-body potential $u({\bf r})$, the corresponding 
Boltzmann weight reads as $w({\bf r}) = \exp[-\beta u({\bf r})]$.
The modified partition function
\begin{equation} \label{partitiongen}
Z_N[w] = \frac{1}{N!} \int_{\Lambda} \prod_{i=1}^N \frac{{\rm d}{\bf r}_i}{
\lambda^3}\, w({\bf r}_i) {\rm e}^{-\beta E(\{ {\bf r}_i\})}
\end{equation} 
is the functional generator for the particle density at point ${\bf r}$: 
\begin{equation}
\rho({\bf r}) = \frac{\delta}{\delta w({\bf r})} \ln Z_N[w] 
\Big\vert_{w({\bf r})=1} .
\end{equation}
With regard of Eqs. (\ref{totalenergy}) and (\ref{deltaE}), it holds that
\begin{eqnarray} 
\ln Z_N[w] & = & \frac{N}{2} \ln \left[ \int_{\Lambda} {\rm d}{\bf r}\, 
w({\bf r}) {\rm e}^{-\kappa\widetilde{z}} \right] \nonumber \\ & & + 
\frac{N}{2} \ln \left[ \int_{\Lambda} {\rm d}{\bf r}\, w({\bf r})
{\rm e}^{-\kappa(\widetilde{d}-\widetilde{z})} \right] . \label{lnZ}
\end{eqnarray}
The functional derivative of this equation with respect to $w({\bf r})$ 
is straightforward: 
\begin{eqnarray}
\frac{\delta}{\delta w({\bf r})} \frac{N}{2} 
\ln \left[ \int_{\Lambda} {\rm d}{\bf r}\, w({\bf r}) 
{\rm e}^{-\kappa\widetilde{z}} \right] \Bigg\vert_{w({\bf r})=1} 
& = & \frac{N {\rm e}^{-\kappa\widetilde{z}}}{
2\int_{\Lambda}{\rm d}{\bf r}\, {\rm e}^{-\kappa\widetilde{z}}} \nonumber \\
& = & \frac{N\kappa {\rm e}^{-\kappa\widetilde{z}}}{2 S \mu
\left( 1-{\rm e}^{-\kappa\widetilde{d}}\right)} \nonumber \\ &  &
\end{eqnarray}
and a similar expression for the second term on the rhs of (\ref{lnZ})
with the substitution $\widetilde{z}\to \widetilde{d}-\widetilde{z}$.
Since $N/(2S\mu) = 2\pi\ell_{\rm B}\sigma^2$, one arrives at the density
profile in the leading SC order,
\begin{equation} \label{rho0}
\widetilde{\rho}(\widetilde{z}) = \frac{\kappa}{1-{\rm e}^{-\kappa\widetilde{d}}}
\left[ {\rm e}^{-\kappa\widetilde{z}} + {\rm e}^{-\kappa(\widetilde{d}-\widetilde{z})} \right] .
\end{equation} 
This formula has the correct reflection 
$\widetilde{z}\to (\widetilde{d}-\widetilde{z})$ symmetry and satisfies
the normalization condition (\ref{normatilde}).

\end{appendix}


\begin{thebibliography}{10}

\bibitem{Levin02}
Y. Levin, Rep. Prog. Phys. {\bf 65}, 1577 (2002).

\bibitem{Andelman06} D. Andelman, in
{\it Soft Condensed Matter Physics in Molecular and Cell Biology},
edited by W.C.K. Poon and D Andelman (Taylor \& Francis, New York, 2006). 

\bibitem{Palberg04} T. Palberg, M. Medebach, N. Garbow, M. Evers, 
A. Barreira Fontecha, H. Reiber, and E. Bartsch, 
J. Phys.: Condens. Matter \textbf{16}, S4039 (2004).

\bibitem{Attard96} Ph. Attard,
Adv. Chem. Phys. \textbf{92}, 1 (1996).

\bibitem{Hansen00} J. P. Hansen and H. L\"owen,
Annu. Rev. Phys. Chem. \textbf{51}, 209 (2000).

\bibitem{Messina09} R. Messina,
J. Phys.: Condens. Matter \textbf{21}, 113102 (2009).

\bibitem{Khan85} A. Khan, B. J\"onsson, and H. Wennerstr\"om,
J. Chem. Phys. \textbf{89}, 5180 (1985).

\bibitem{Kjellander88} R. Kjellander, S. Mar\v{c}elja, and J. P. Quirk,
J. Colloid Interface Sci. \textbf{126}, 194 (1988).

\bibitem{Bloomfield91} V. A. Bloomfield, 
Biopolymers \textbf{31}, 1471 (1991).

\bibitem{Rau92} D. C. Rau and A. Pargesian, 
Biophys. J. \textbf{61}, 246 (1992); ibid. \textbf{61}, 260 (1992).

\bibitem{Kekicheff93} P. K\'ekicheff, S. Mar\v{c}elja, T. J. Senden, and 
V. E. Shubin, J. Chem. Phys. \textbf{99}, 6098 (1993).


\bibitem{Komorowski18}
K. Komorowski, A. Salditt, Y. Xu, H. Yavuz, M. Brennich, R. Jahnand and T. Salditt,
Biophys. Journal {\bf 114} 1908, (2018).

\bibitem{Mukina19}
T. Mukhina, A. Hemmerle, V. Rondelli, Y. Gerelli, G. Fragneto,
J. Daillant and T. Charitat,
J. Phys. Chem. Lett. {\bf 10}, 7195 (2019).

\bibitem{Fink19}
L. Fink, A. Steiner, O. Szekely, P. Szekely and U. Raviv,
Langmuir {\bf 35}, 9694 (2019).

\bibitem{Komorowski20}
K. Komorowski, J. Schaeper, M. Sztucki, L. Sharpnack,
G. Brehm, S. K\"oster and T. Salditt, Soft Matter {\bf 16}, 4142 (2020).

\bibitem{Guldbrand84} L. Guldbrand, B. J\"onsson, H. Wennerstr\"om, and H. Linse,
J. Chem. Phys. \textbf{80}, 2221 (1984).

\bibitem{Kjellander84} R. Kjellander and S. Mar\v{c}elja,
Chem. Phys. Lett. \textbf{112}, 49 (1984).

\bibitem{Bratko86} D. Bratko, B. J\"onsson, and H. Wennerstr\"om,
Chem. Phys. Lett. \textbf{128}, 449 (1986).

\bibitem{Gronbech97} N. Gr{\o}nbech-Jensen, R. J. Mashl, R. F. Bruinsma,
and W. M. Gelbart, Phys. Rev. Lett. \textbf{78}, 2477 (1997). 

\bibitem{rque1}
Although the dielectric constant of colloids in general differs
from that of the surrounding medium (such as water) in which the counterions are
immersed, one usually uses a simplified model with no such discontinuity and thus no dielectric images. 




\remma{
\bibitem{Gebbie13} M.~A. Gebbie, M. Valtiner, X. Banquy, E.~T. Fox, W.~A. Henderson, and J.~N. Israelachvili,
Proc. Natl. Acad. Sci. U. S. A.  \textbf{110}, 9674 (2013).

\bibitem{Valmacco15} V. Valmacco , G. Trefalt , P. Maroni and M. Borkovec,
Phys. Chem. Chem. Phys. \textbf{17}, 16553 (2015).

\bibitem{Labbez06} C. Labbez,  B. J\"onsson, I. Pochard, A. Nonat, and Bernard Cabane,
J. Phys. Chem. B  \textbf{110}, 9219 (2006). 

\bibitem{Smith16} A.~M. Smith, A.~A. Lee, and S. Perkin: 
J. Phys. Chem. Lett. \textbf{7}, 2157 (2016).

\bibitem{Jing15} Y. Jing, V. Jadhao, J.~W. Zwanikken, and M. Olvera de la Cruz;
J. Chem. Phys. \textbf{143}, 194508 (2015).
}

\bibitem{Edwards62} S. F. Edwards and A. Lenard,
J. Math. Phys. \textbf{3}, 778 (1962).

\bibitem{Attard88} Ph. Attard, D. J. Mitchell, and B. W. Ninham,
J. Chem Phys. \textbf{88}, 4987 (1988); \textbf{89}, 4358 (1988);
R. Podgornik, J. Phys. A \textbf{23}, 275 (1990);
R. R. Netz and H. Orland, Eur. Phys. J. E \textbf{1}, 203 (2000). 

\bibitem{Moreira00} A. G. Moreira and R. R. Netz:
Europhys. Lett. \textbf{52}, 705 (2000); 
Phys. Rev. Lett. \textbf{87}, 078301 (2001).

\bibitem{Netz01} R. R. Netz:
Eur. Phys. J. E \textbf{5}, 557 (2001).

\bibitem{Moreira02} A. G. Moreira and R. R. Netz:
Eur. Phys. J. E \textbf{8}, 33 (2002).

\bibitem{Kanduc07} M. Kandu\v{c} and R. Podgornik,
Eur. Phys. J. E \textbf{23}, 265 (2007);
Y. S. Jho, M. Kandu\v{c}, A. Naji, R. Podgornik, M. W. Kim, and P. A. Pincus,
Phys. Rev. Lett. \textbf{101}, 188101 (2008).

\bibitem{Rouzina96} I. Rouzina and V. A. Bloomfield,
J. Phys. Chem. \textbf{100}, 9977 (1996).

\bibitem{Shklovskii99} B. I. Shklovskii,
Phys. Rev. E \textbf{60}, 5802 (1999);
Phys. Rev. Lett. \textbf{82}, 3268 (1999).

\bibitem{Perel99} V. I. Perel and B. I. Shklovskii, 
Physica A \textbf{274}, 446 (1999);

\bibitem{Earnshaw1842} S. Earnshaw,
Trans. Cambridge Philos. Soc. \textbf{7}, 97 (1842).

\bibitem{Falko94} V.I. Falko,
Phys. Rev. B \textbf{49}, 7774 (1994).

\bibitem{Esfarjani95} K. Esfarjani and Y. Kawazoe,
J. Phys.: Condens. Matter \textbf{7} 7217 (1995).

\bibitem{Goldoni96} G. Goldoni and F. M. Peeters,
Phys. Rev. B \textbf{53}, 4591 (1996).

\bibitem{Schweigert99} I. V. Schweigert, V. A. Schweigert, and F. M. Peeters, 
Phys. Rev. Lett. \textbf{82}, 5293 (1999); Phys. Rev. B \textbf{60}, 14 665
(1999).

\bibitem{Weis01} J. J. Weis, D. Levesque, and S. Jorge,
Phys. Rev. B \textbf{63}, 045308 (2001).

\bibitem{Messina03} R. Messina and H. L\"owen, 
Phys. Rev. Lett. \textbf{91}, 146101 (2003); 
E. C. O\v{g}uz, R. Messina, and H. L\"owen, 
Europhys. Lett. \textbf{86}, 28002 (2009).

\bibitem{Lobaskin07} V. Lobaskin and R. R. Netz,
Europhys. Lett. \textbf{77}, 38003 (2007).

\bibitem{Samaj12a} L. \v{S}amaj and E. Trizac,
Europhys. Lett. \textbf{98}, 36004 (2012); 
Phys. Rev. B \textbf{85}, 205131 (2012).

\bibitem{Misra} R. D. Misra, Math. Proc. Cambridge Philos. Soc.
\textbf{36}, 173 (1940); 
M. Born and R. D. Misra, Math. Proc. Cambridge Philos. Soc.
\textbf{36}, 466 (1940). 

\bibitem{Samaj11} L. \v{S}amaj and E. Trizac,
Phys. Rev. Lett. \textbf{106}, 078301 (2011);
Phys. Rev. E \textbf{84}, 041401 (2011).

\bibitem{Varenna}
E. Trizac and L. \v{S}amaj, in Proceedings of the International
School of Physics Enrico Fermi, edited by C. Bechinger, F.
Sciortino and P. Ziherl, Vol. 184 (2013), p. 61.

\bibitem{Samaj18} L. \v{S}amaj, M. Trulsson, and E. Trizac, 
Soft Matter \textbf{14}, 4040 (2018).

\bibitem{Chen06} Y. G. Chen and J. D. Weeks,
Proc. Natl. Acad. Sci. U. S. A. \textbf{103}, 7560 (2006);
J. M. Rodgers, C. Kaur, Y. G. Chen, and J. D. Weeks
Phys. Rev. Lett. \textbf{97}, 097801 (2006).

\bibitem{Nordholm84} S. Nordholm,
Chem. Phys. Lett. \textbf{105}, 302 (1984).

\bibitem{Santangelo06} C. D. Santangelo,
Phys. Rev. E \textbf{73}, 041512 (2006).

\bibitem{Hatlo10} M. M. Hatlo and L. Lue,
EPL \textbf{89}, 25002 (2010).

\bibitem{Bakhshandeh11} A. Bakhshandeh, A. P. dos Santos, and Y. Levin,
Phys. Rev. Lett. \textbf{107}, 107801 (2011).

\bibitem{Palaia18} I. Palaia, M. Trulsson, L. \v{S}amaj, and E. Trizac,
Mol. Phys. \textbf{116}, 3134 (2018). 









\bibitem{Lek11} H. N. Lekkerkerker and R. Tuinier,
{\it Colloids and the Depletion Interaction},
Lecture Notes in Physics 833 (Springer, 2011).

\bibitem{Schmidt97} M. Schmidt and H. L\"owen,
Phys. Rev. E \textbf{55}, 7228 (1997).

\bibitem{Borukhov97} I. Borukhov, D. Andelman, and H. Orland,
Phys. Rev. Lett. \textbf{79}, 435 (1997). 

\bibitem{Patra02} C. N. Patra and S. K. Ghosh,
J. Chem. Phys.\textbf{117}, 8938 (2002). 

\bibitem{Kilic07} M. S. Kilic, M.Z. Bazant, and A. Ajdari,
Phys. Rev. E \textbf{75}, 021502 (2007).

\bibitem{Valleau91} J. P. Valleau, R. Ivkov, and G. M. Torrie, 
J. Chem. Phys. \textbf{95}, 520 (1991).

\bibitem{Kjellander92} R. Kjellander, T. Akesson, B. J\"{o}nsson, and 
S. Mar\v{c}elja, J. Chem. Phys. \textbf{97}, 1424 (1992).

\bibitem{Zelko10} J. Zelko, A. Igli\v{c}, V. Kralj-Igli\v{c}, and 
P. B. S. Kumar,
J. Chem. Phys. \textbf{133}, 204901 (2010).

\bibitem{Grimes79} C. C. Grimes and G. Adams,
Phys. Rev. Lett. \textbf{42}, 795 (1979). 

\bibitem{Morf79} R. H. Morf,
Phys. Rev. Lett. \textbf{43}, 931 (1979).

\bibitem{contact} D. Henderson and L. Blum, J. Chem. Phys. \textbf{69}, 
5441 (1978); D. Henderson, L. Blum, and J. L. Lebowitz, 
J. Electroanal. Chem. \textbf{102}, 315 (1979);
S. L. Carnie, D.Y.C. Chan, J. Chem. Phys. \textbf{74}, 1293 (1981);
H. Wennerstr\"om, B. J\"onsson, and P. Linse,
J. Chem. Phys. \textbf{76}, 4665 (1982).

\bibitem{Feynman98} R. P. Feynman,
Statistical Mechanics: A Set of Lectures
(Westview Press, 1998).

\bibitem{A}
I.-C. Yeh and M. L. Berkowitz, J. Chem. Phys. {\bf 111}, 3155 (1999).

\bibitem{B}
M. Mazars, J.-M. Caillol, J.-J. Weis and D. Levesque, Condens. Matter Phys. {\bf 4}, 697 (2001).

\bibitem{Ivan}
I. Palaia, Charged systems in, out of, and driven to equilibrium: from nanocapacitors to cement, PhD, Paris-Saclay University (2019). 

\bibitem{Palaia20}
I. Palaia, A. Goyal, E. del Gado. L. \v{S}amaj and E. Trizac, to be published.

\bibitem{rque100}
In Fig. \ref{fig:Structures}-a) indeed, the triangular lattice corresponds to having $\Delta=\sqrt{3}$.

\bibitem{Schlaich19}
A. Schlaich, A.P. dos Santos, R.R. Netz, 
Langmuir {\bf 35}, 551 (2019).

\bibitem{Gradshteyn} I. S. Gradshteyn and I. M. Ryzhik,
Table of Integrals, Series, and Products, 6th ed.
(Academic, London, 2000).

\bibitem{Travenec15} I. Trav\v{e}nec and L. \v{S}amaj,
Phys. Rev. E \textbf{92}, 022306 (2015). 

\bibitem{Paillusson11} 
\remma{F. Paillusson and E. Trizac,
Phys. Rev. E \textbf{84}, 011407 (2011). }



\end{thebibliography}
\end{document}